\documentclass[11pt, a4paper]{article}
\usepackage{jheparxiv}
\usepackage[utf8]{inputenc}
\usepackage{amsmath}
\usepackage{amsfonts}
\usepackage{amssymb}
\usepackage{latexsym}
\usepackage{mathrsfs}
\usepackage{braket}		%Bra Ket Notation
\usepackage{graphicx}
\usepackage{color}
\usepackage{xcolor}
\usepackage{slashed}
\usepackage{twistor}
\usepackage[all]{xy}
\usepackage{tikz-cd}
\usepackage{mathtools}
\usepackage{tikz-feynman}

\newcommand{\bb}{\mathbf{b}}

\renewcommand{\d}{\mathrm{d}}

\newcommand{\g}{\mathfrak{g}}

\newcommand{\bigma}{\boldsymbol{\sigma}}

\def\mL{\mathcal{L}}

\def\mM{\mathcal{M}}

\def\mM{\mathcal{M}}

\def\mA{\mathcal{A}}

\def\zb{\bar{z}}

\def\pa{\partial}

\renewcommand{\[}{\begin{equation}\begin{aligned}}
\renewcommand{\]}{\end{aligned}\end{equation}}

\def\g5{\gamma_5}

\def\mt{\tilde{m}}

\def\Qt{\tilde{Q}}
\def\Tt{\tilde{T}}

\def\b[#1]{\bold{#1}}
\def\bb[#1]{\overline{\bold{#1}}}
\def\bs[#1,#2]{\bold{#1}_{#2}}
\def\bbs[#1,#2]{\overline{\bold{#1}}_{#2}}

\def\s2{\sigma_2}
\def\ep{\epsilon}

\def\gammaflat{ \gamma_{z\zb}}
\def\gammaflatt{ \gamma^{z\zb}}

\def\Tsoft{T_{\text{soft}}}
\def\Thard{T_{\text{hard}}}

\def\paz{\pa_z}

\def\pazb{\pa_{\zb}}

\def\ketd[#1]{\ket{#1}_{\text{dressed}}}
\def\brad[#1]{\bra{#1}_{\text{dressed}}}
\def\ketas[#1]{\ket{#1}_{\text{Asymptotic}}}
\def\braas[#1]{\bra{#1}_{\text{Asymptotic}}}
\def\Qsoft{Q_{\text{soft}}}
\def\Qhard{Q_{\text{hard}}}

\title{Classical Double Copy at Null Infinity}

\author[a]{Tim Adamo}
\author[b]{\& Uri Kol}

\affiliation[a]{School of Mathematics and Maxwell Institute for Mathematical Sciences \\
        University of Edinburgh, EH9 3FD, UK}

\affiliation[b]{Center for Cosmology and Particle Physics, Department of Physics \\
		New York University, 726 Broadway, New York, NY 10003, USA}

\emailAdd{t.adamo@ed.ac.uk}
\emailAdd{urikol@gmail.com}

\abstract{We give two double copy prescriptions which construct asymptotically flat solutions in gravity from asymptotically flat gauge fields. The first prescription applies to radiative fields, which are non-linear vacuum solutions determined by characteristic data at null infinity. For any two such radiative gauge fields (linear or non-linear), the characteristic data of a radiative metric, dilaton and axion is constructed by a simple `squaring' procedure, giving a classical double copy at the level of radiation fields. We demonstrate the procedure with several examples where the characteristic data can be explicitly integrated; for linear fields this also sheds light on the twistorial description of Weyl double copy. Our second prescription applies to all asymptotically flat fields at the level of their asymptotic equations of motion: we give a map between any solution of the asymptotic Maxwell equations and any solution of the asymptotic Einstein equations at null infinity. This also extends to the asymptotic charges and their duals, preserves the soft and hard sectors between gauge theory and gravity, and is related to the usual notion of double copy in scattering amplitudes.}

\begin{document}

\maketitle

\section{Introduction}

Double copy is now an extremely well-studied concept in the context of scattering amplitudes~\cite{Bern:2019prr,Borsten:2020bgv}: with its origins in the KLT relations of string theory~\cite{Kawai:1985xq}, double copy can be roughly characterized by saying that amplitudes in a gravitational theory can be obtained `for free' from those of a non-gravitational (usually gauge) theory, provided the amplitudes in the latter are expressed in a suitable (color-kinematics~\cite{Bern:2008qj,Bern:2010ue}) representation. In its simplest incarnation, this relates amplitudes in pure Yang-Mills theory to amplitudes in `NS-NS' (or $\cN=0$) supergravity, composed of a metric, dilaton and Kalb-Ramond $B$-field. The manifold successes of double copy in the context of amplitudes -- and the fact that amplitudes arise from recursively constructed solutions to the equations of motion (cf., \cite{Arefeva:1974jv,Jevicki:1987ax,Rosly:1996vr,Selivanov:1997aq,Mizera:2018jbh,Cho:2021nim}) -- raises the question: is it possible that some notion of double copy could extend to \emph{classical, non-linear} solutions to the underlying field theories on either side of the correspondence?

There are, of course, many different things that one could mean when asking this question. For instance, one could consider the perturbative construction of a gauge field and ask if this can be mapped, order-by-order, into the perturbative construction of a metric. Such a perturbative classical double copy is indeed possible~\cite{Goldberger:2016iau,Luna:2016hge,Goldberger:2017frp,Goldberger:2017vcg,Luna:2017dtq,Chester:2017vcz,CarrilloGonzalez:2018ejf,Plefka:2018dpa,Prabhu:2020avf}, although it becomes fairly complicated already at next-to-leading order~\cite{Shen:2018ebu}. Alternatively, one could try to construct a classical double copy which operates at the level of \emph{exact} solutions, sending a solution of the Yang-Mills equations to a solution of the Einstein equations or the NS-NS sector of type II supergravity.

In~\cite{Monteiro:2014cda,Luna:2015paa}, the first realization of such an exact classical double copy was given for highly symmetric solutions in gravity that admit Kerr-Schild coordinates (and are hence algebraically special). This `Kerr-Schild double copy' has been extensively developed and generalized in recent years, and is closely linked to the scattering amplitude interpretation of double copy\footnote{Interestingly, a linearised version of Kerr-Schild double copy was first written down in the context of higher-spin gravity~\cite{Didenko:2008va}, where it played a crucial role in obtaining higher-spin black hole solutions~\cite{Didenko:2009td}.}. However, there are several clear limitations to this notion of classical double copy. Firstly, the existence of Kerr-Schild (or generalized Kerr-Schild) coordinates is highly constraining: although many exact solutions in GR admit Kerr-Schild coordinates (including all black holes in the Kerr-Newman family of electrovacuum solutions), these are a set of measure zero inside the space of solutions to the Einstein equations. Furthermore, Kerr-Schild double copy is practically used as a map from gravitational to gauge theory solutions: one starts with a metric in Kerr-Schild coordinates and identifies a Maxwell field. This `single copy' runs counterintuitive to the usual sense of double copy\footnote{It has recently been shown that a generalization of the Kerr-Schild double copy, known as the Weyl double copy~\cite{Luna:2018dpt}, can be understood in a truly double copy sense by mapping onto scattering amplitudes -- at least at the level of 3-point amplitudes~\cite{Monteiro:2020plf}.}.

\medskip

Is there a notion of classical double copy which is more general and truly maps gauge theory solutions into gravitational solutions? In this paper, we provide two complementary answers to this question by considering \emph{asymptotically flat} solutions in four space-time dimensions. These solutions admit a well-defined conformal compactification in the sense of Penrose~\cite{Penrose:1962ij,Penrose:1964ge}, where a conformal rescaling endows the space-time (Minkowski space for the gauge fields, or the space-time itself for gravity) with a null conformal boundary, $\scri$. This \emph{null infinity} is composed of two distinct pieces, a future $\scri^+$ and past $\scri^-$, each with topology $\scri^{\pm}\cong\R\times S^2$.

Exploiting the structure of null infinity available, we present two complementary notions of classical double copy in this paper. The first applies to \emph{radiative} fields, which are asymptotically flat gauge and gravitational fields for which $\scri^+$ is a good characteristic surface for the field equations\footnote{Note that $\scri^+$ is not a good final data surface for all space-times~\cite{Geroch:1978us}; this condition defines the class of radiative space-times. Furthermore, the existence of a smooth $\scri^+\cong\R\times S^2$ is not a generic condition for solutions of the Einstein equations. Indeed, fairly simple incoming matter will generate non-smooth $\scri^+$~\cite{Kehrberger:2021uvf}. However, such examples lie outside of the class of radiative space-times, being generated by matter sources in the asymptotic past.}. Radiative fields are source-free and determined by a freely-specified function (of appropriate spin and conformal weight) on $\scri^+$, which serves as the characteristic data. Our double copy prescription takes any radiative Yang-Mills field and defines a radiative NS-NS gravitational field (i.e., a metric, dilaton and $B$-field) at the level of this characteristic data.

On the one hand, this construction is fairly general, applying to \emph{any} radiative gauge field, without assumptions of linearity or algebraic speciality. On the other hand, it has several shortcomings: in general it is not possible to reconstruct the `bulk' gauge and gravitational fields explicitly from their characteristic data; one only knows that such bulk solutions exist (at least locally) and are uniquely determined by the data~\cite{Sachs:1962zzb,Friedrich:1986rb}. Furthermore, it is easy to see by considering the special case of linear fields (where the bulk field is reconstructed explicitly via a Kirchhoff-d'Adh\'emar integral formula~\cite{Penrose:1980yx}) that this notion of double copy is not directly related to the version arising in the study of scattering amplitudes. Nevertheless, this prescription has some features and motivations in common with other studies of classical double copy, particularly~\cite{Elor:2020nqe,Easson:2020esh,Casali:2020uvr,Chacon:2020fmr,Pasterski:2020pdk,Campiglia:2021srh,Farnsworth:2021wvs}.

\medskip

The second version of double copy at null infinity we introduce acts on \emph{any} asymptotically flat field, at the level of the asymptotic equations of motion on $\scri^+$. This version of double copy is motivated by the universality of the 3-point coupling of matter to photons and gravitons in the low energy limit, as manifested by the factorization of any scattering amplitude according to the soft theorem
\[
\mA_{k} \sim \frac{-\im }{p\cdot k } \times	( p \cdot \ep)^h
\times \mA\, ,
\]
with $h=(1,2)$ for photons and gravitons, respectively. Here $\mA_{k}$ is any amplitude with a soft external photon/graviton leg of momentum $k$, $\mA$ is the same amplitude without the soft leg, $-\im/(p\cdot k)$ is the extra matter propagator and
\[\label{threePoint}
\mM_3 = ( p \cdot \ep)^h
\]
is the 3-point coupling.
%$\ep_{\mu}$ is the polarization vector of the photon and for the transverse-traceless components of the graviton we can decompose the polarization tensor as $\ep_{\mu\nu}=\ep_{\mu}\ep_{\nu}$.
This famous result holds, in the low energy limit, for matter fields of any spin~\cite{Weinberg:1965nx}. Clearly the 3-point amplitude \eqref{threePoint} obeys a double copy structure - the gravitational amplitude is double the electromagnetic one.
The universality of this 3-point amplitude suggests that in the low energy limit, or equivalently at large distances, any classical solution of Maxwell's theory can be mapped into a classical solution of Einstein's gravity.

Aligned with this expectation, we give an explicit map which takes any solution of the Maxwell equations and produces a solution of the Einstein equations order-by-order in the peeling expansion near $\scri^+$. This map extends to asymptotic charges, with the asymptotic electric and magnetic charges of Maxwell theory being mapped into the supertranslation and dual supertranslation charges of gravity. The soft and hard sectors of each charge  are also preserved by this map. Furthermore, we show explicitly how this notion of classical double copy is directly related to the double copy structure of the 3-point amplitude \eqref{threePoint}.

The two double copy prescriptions that we discuss in this paper are both defined using the structure of the theories at null infinity.
However, while the first one preserves the radiative structure of the fields, the second prescription does not (since it mixes the radiative degrees of freedom with the Coulomb components of the fields). We therefore view these two prescription as complementary to each other.

\medskip

The paper is organized as follows: Section~\ref{DCrad} sets out a classical double copy prescription at null infinity for radiative fields. This is most transparent in a `homogeneous' description of null infinity, which we review. Section~\ref{RadEx} provides several examples of this prescription in action (for linear fields and known exact solutions), and also comments on its relationship with the recently proposed `twistorial' double copy~\cite{White:2020sfn,Chacon:2021wbr,Chacon:2021hfe}. 
In Section~\ref{AsyClassDC} we set out a classical double copy prescription for any asymptomatically flat gauge and gravitational fields, expanded around null infinity.
Section~\ref{Amps} describes how this prescription is related to the usual notion of double copy from scattering amplitudes.

\medskip

\paragraph{Note added:} While writing up this work we became aware of~\cite{Godazgar:2021}, which has similar motivations to the results presented here.

%%%%%%%%%%%%%%%%%%%%%%%%%%
%%%%%%%%%%%%%%%%%%%%%%%%%%

\section{Double Copy for Radiative Fields}
\label{DCrad}

A \emph{radiative} field, of any spin, is an asymptotically flat solution to the non-linear vacuum equations which is completely determined by characteristic data at past or future null infinity, $\scri^{\pm}$. This characteristic data can be viewed as specifying incoming or outgoing radiation data, which is in turn integrated to uniquely determine the radiative field in the bulk. For fixed spin, the characteristic data is a function of fixed spin and conformal weight on $\scri^{\pm}$, but besides some elementary smoothness and regularity assumptions, can be freely specified. That this is also true for gravity is particularly interesting, since the existence and smoothness of the null conformal boundary itself depends on the properties of the physical (i.e., non-conformally-rescaled) metric~\cite{Sachs:1962zzb,Friedrich:1986rb}. Thus, the space of radiative fields is extremely large, even if explicit non-linear examples are difficult to construct.

In this section, we give a double copy prescription for radiative gauge fields at the level of their characteristic data. We will always work on future null infinity, $\scri^+$, where the characteristic data specifies outgoing radiation; it is trivial to transform all of what follows to $\scri^-$ and incoming radiation data. Given any two radiative gauge fields in Minkowski space with the same gauge group, we show that characteristic data for a radiative metric and two radiative scalars is constructed by simply multiplying the characteristic data for the gauge fields and tracing over the gauge group. Identifying these two scalars as a dilaton and axion, this yields precisely the field content of a NS-NS gravitational field, as predicted by the double copy.

%%%%%%%%%%%%%%%%%%%%%%%%%%

\subsection{Homogeneous geometry of $\scri^+$}

In retarded Bondi coordinates $(u,r,z,\bar{z})$ and Bondi-Sachs gauge, an asymptotically flat space-time admits a large-$r$ expansion of the form~\cite{Bondi:1962px,Sachs:1962wk,Madler:2016xju}:
\begin{multline}\label{BS1}
\d s^2=-\left(1-2\,\frac{m_{B}(u,z,\bar{z})}{r}\right) \d u^2 - 2\,\d u\,\d r +\bar{\eth}\bigma^0(u,z,\bar{z})\,\d u\,\d z+\eth\bar{\bigma}^0(u,z,\bar{z})\,\d u\,\d\bar{z}\\
+\frac{2\,r^2}{(1+|z|^2)^2}\left(\d z\,\d\bar{z}+\frac{\bigma^0(u,z,\bar{z})}{r}\,\d z^2+ \frac{\bar{\bigma}^0(u,z,\bar{z})}{r}\,\d \bar{z}^2\right)+O(r^{-2})\,,
\end{multline}
where $m_{B}$ is the Bondi mass aspect (spin weight 0 and conformal weight $-3$), $\eth$ is the spin-weighted covariant derivative on the sphere~\cite{Goldberg:1966uu} and $\bigma^0$, $\bar{\bigma}^0$ are the optical parameters encoding the asymptotic shear of constant-$u$ null hypersurfaces~\cite{Jordan:1961,Sachs:1961zz}, with conformal weight $-1$ and spin weights $-2$ and $2$, respectively\footnote{Our conventions for spin and conformal weight, which are chosen to match those in the amplitudes literature, have the slightly unfortunate consequence of making $\eth$ a spin-lowering operator and $\bar{\eth}$ a spin-raising operator.}. Equivalently, $\bigma^0$ can be viewed as the leading coefficient in the large-$r$ expansion of the spin coefficient $\sigma$ in the Newman-Penrose formalism~\cite{Newman:1961qr,Newman:2009,Adamo:2009vu}\footnote{Our notation is chosen to mirror that of the Newman-Penrose formalism as closely as possible, but comparison with Bondi-style notation is straightforward by taking $\frac{2\bigma^0}{(1+|z|^2)^2}\leftrightarrow C_{zz}$, $\bar{\eth}\bigma^0\leftrightarrow D^{z}C_{zz}$, etc.}.

Here, $(z,\bar{z})$ are complex stereographic coordinates on the sphere, related to the usual angular variables $(\theta,\phi)$ by $z=\e^{\im\phi}\cot(\theta/2)$, so these coordinates break manifest Lorentz invariance. While this shortcoming is not too serious, it is sometimes advantageous to work in a formalism with manifest Lorentz invariance by lifting the retarded Bondi coordinates to a \emph{homogeneous}, or projective, formalism (cf., \cite{Eastwood:1982,Sparling:1990,Adamo:2014yya,Geyer:2014lca,Adamo:2015fwa,Adamo:2020yzi}). This is achieved by working with coordinates $(u,r,\lambda_{\alpha},\bar{\lambda}_{\dot\alpha})$, where $\alpha=0,1$, $\dot\alpha=\dot{0},\dot{1}$ are $\SL(2,\C)$ Weyl spinors of opposite chirality and these coordinates are identified up to overall rescalings
\be\label{projcoord}
(u,r,\lambda_\alpha,\bar\lambda_{\dot\alpha})\sim(|b|^2 u,\,|b|^{-2}r,\,b\lambda_{\alpha},\,\bar{b}\bar{\lambda}_{\dot\alpha})\,,
\ee
for any non-zero complex number $b$. The null vector $\lambda^{\alpha}\bar{\lambda}^{\dot\alpha}$ (which is invariant under these scalings) is tangent to the outgoing geodesics of the constant-$u$ hypersurfaces (i.e., $\lambda^{\alpha}\bar{\lambda}^{\dot\alpha}\partial_{\alpha\dot\alpha}=\partial_r$).

It is, of course, straightforward to pass between this homogeneous coordinate system and the standard retarded Bondi coordinates by working on an affine patch; for the coordinates on the sphere, this affine patch is defined by
\be\label{affine1}
\lambda_{\alpha}=\frac{2^{1/4}}{\sqrt{1+|z|^2}}\,(-z,\,1)\,, \qquad \bar{\lambda}_{\dot\alpha}=\frac{2^{1/4}}{\sqrt{1+|z|^2}}\,(-\bar{z},\,1)\,.
\ee
The remaining non-homogeneous Bondi coordinates $(u,r)$ are recovered by choosing a future pointing time-like vector $t^{\alpha\dot\alpha}$, normalised so that $t^2=1$, and defining
\be\label{hat}
\hat{\lambda}_{\alpha}:=t_{\alpha\dot\alpha}\,\bar{\lambda}^{\dot\alpha}\,.
\ee
This choice of $t^{\alpha\dot\alpha}$ is equivalent to fixing a conformal scale on the sphere (as we will see later); a standard choice is $t^{\alpha\dot\alpha}=\mathrm{diag}(\frac{1}{\sqrt{2}},\frac{1}{\sqrt{2}})$ for which $\la\lambda\,\hat{\lambda}\ra=1$. Standard Bondi coordinates are then encoded in spinor form by
\be\label{affine2}
x^{\alpha\dot\alpha}=\frac{u}{\la\lambda\hat{\lambda}\ra}\,t^{\alpha\dot\alpha}+r\,\lambda^{\alpha}\,\bar{\lambda}^{\dot\alpha}\,,
\ee
evaluated on the affine patch \eqref{affine1}. These variables also enable a spinor description of the standard Bondi null tetrad, $\{l^a, n^a, m^a,\bar{m}^a\}$:
\be\label{Bonditetrad}
l^{\alpha\dot\alpha}=\lambda^{\alpha}\,\bar{\lambda}^{\dot\alpha}\,, \qquad n^{\alpha\dot\alpha}=\frac{\hat{\lambda}^{\alpha}\,\bar{\hat{\lambda}}^{\dot\alpha}}{\la\lambda\,\hat{\lambda}\ra^2}\,, \qquad m^{\alpha\dot\alpha}=\frac{\hat{\lambda}^{\alpha}\,\bar{\lambda}^{\dot\alpha}}{\la\lambda\,\hat{\lambda}\ra}\,,
\ee
which obey $l\cdot n=1=-m\cdot \bar{m}$, with all other inner products vanishing.

In this homogeneous formalism, the asymptotic Bondi-Sachs expansion \eqref{BS1} becomes
\begin{multline}\label{BS2}
\d s^2=-\left(\frac{1}{\la\lambda\,\hat\lambda\ra^{2}}-2\,\frac{m_{B}(u,\lambda,\bar{\lambda})}{r}\right) \d u^2 - 2\,\d u\,\d r +\d u\left(\bar{\eth}\bigma^0(u,\lambda,\bar{\lambda})\,\D\lambda+\eth\bar{\bigma}^0(u,\lambda,\bar{\lambda})\,\D\bar{\lambda}\right)\\
+r^2\,\left(\D \lambda\,\D\bar{\lambda}-\frac{\bigma^0(u,\lambda,\bar{\lambda})}{r}\,\D\lambda^2- \frac{\bar{\bigma}^0(u,\lambda,\bar{\lambda})}{r}\,\D\bar{\lambda}^2\right)+O(r^{-2})\,,
\end{multline}
where $\D\lambda:=\la\lambda\,\d\lambda\ra=\lambda^{\alpha}\d\lambda_{\alpha}$, $\D\bar{\lambda}:=[\bar{\lambda}\,\d\bar{\lambda}]=\bar{\lambda}^{\dot\alpha}\d\bar{\lambda}_{\dot\alpha}$ and the $\eth$-operator in the homogeneous formalism is defined by~\cite{Eastwood:1982}
\be\label{eth}
\eth f(\lambda, \bar{\lambda})=\frac{\hat{\lambda}_{\alpha}}{\la\lambda\,\hat{\lambda}\ra}\,\frac{\partial f}{\partial\lambda_{\alpha}}\,,
\ee
where $f$ is homogeneous of any degree in $(\lambda,\bar{\lambda})$; $\bar{\eth}$ is defined by complex conjugation.

The requirement that the line element is weightless under the rescalings \eqref{projcoord} imposes scaling properties on the Bondi mass aspect and the shear optical scalars:
\be\label{MAOscal}
m_{B}(|b|^2u,b\lambda,\bar{b}\bar{\lambda})=|b|^{-3}\,m_{B}(u,\lambda,\bar{\lambda})\,, \qquad \bigma^0(|b|^2u,b\lambda,\bar{b}\lambda)=b^{-3}\,\bar{b}\,\bigma^0(u,\lambda,\bar{\lambda})\,,
\ee
with the scaling of $\bar{\bigma}^0$ defined by complex conjugation. More generally, any function $f$ which transforms as $f\to b^{p}\bar{b}^{q}f$ under \eqref{projcoord} should be viewed as a section of a line bundle $\cO(p,q)$. Thus, $m_B$, $\bigma^0$ and $\bar{\bigma}^0$ are sections of $\cO(-3,-3)$, $\cO(-3,1)$ and $\cO(1,-3)$, respectively.

Given any section of $\cO(p,q)$, the information encoded in $p,q\in\Z$ is equivalent to the usual labels of spin and conformal weight through $s=\frac{p-q}{2}$ and $w=\frac{p+q}{2}$~\cite{Eastwood:1982}. Thus, one easily recovers the appropriate spin and conformal weights of the Bondi mass aspect and asymptotic shear parameters from their scaling behaviour in the homogeneous formalism.

\medskip

Any solution of the Einstein equations admitting a Bondi-Sachs expansion \eqref{BS1}, \eqref{BS2} is asymptotically flat in the sense of Penrose~\cite{Penrose:1962ij,Penrose:1964ge,Newman:1981,Frauendiener:2000mk}. In particular, rescaling the metric by a conformal factor $\Omega:=r^{-1}$ one obtains
\begin{multline}\label{ccBS}
\d \hat{s}^2:=\Omega^2\,\d s^2=-\frac{\Omega^2}{\la\lambda\,\hat{\lambda}\ra^2}\,\d u^2+2\,\d u\,\d\Omega+\D\lambda\,\D\lambda +\Omega^2\left(\bar{\eth}\bigma^0\,\d u\,\D\lambda+\eth\bar{\bigma}^0\,\d u\,\D\bar{\lambda}\right)\\
-\Omega\left(\bigma^0\,\D\lambda^2+\bar{\bigma}^0\,\D\bar{\lambda}^2\right)+2\,\Omega^3\,m_{B}\,\d u^2+O(\Omega^4)\,.
\end{multline}
The future conformal boundary $\scri^+$ is defined by the $\Omega\to0$ (or $r\to\infty$) limit, whence the conformally rescaled metric \eqref{ccBS} becomes the degenerate metric on the $(u,\lambda,\bar{\lambda})$ null cone:
\be\label{scrimet}
\d\hat{s}^2|_{\scri^+}=0\times\d u^2+\frac{2\,\d z\,\d\bar{z}}{(1+|z|^2)^2}=0\times \d u^2+\D\lambda\,\D\bar{\lambda}\,,
\ee
where the homogeneous coordinates on $\scri^+$ are identified up to rescalings $(|b|^2u,b\lambda,\bar{b}\bar{\lambda})\sim (u,\lambda,\bar{\lambda})$. Thus, in the homogeneous formalism $\scri^+$ is the total space of the line bundle $\cO_{\R}(1,1)\rightarrow\CP^1$, where sections of $\cO_{\R}(1,1)$ are real-valued and obey $f(b\lambda,\bar{b}\bar{\lambda})=|b|^2 f(\lambda,\bar{\lambda})$, the $(\lambda_{\alpha},\bar{\lambda}_{\dot\alpha})$ serve as homogeneous coordinates on $\CP^1\cong S^2$ and $u$ is the fibre coordinate. Note that the conformally rescaled line element \eqref{scrimet} on $\scri^+$ takes values in $\cO(2,2)$, so must be normalised by a choice of conformal factor on the constant-$u$ cross sections. The standard choice
\be\label{scrimet2}
\d s^2|_{\scri^+}=0\times \d u^2+\frac{\D\lambda\,\D\bar{\lambda}}{\la\lambda\,\hat{\lambda}\ra^2}\,,
\ee
endows constant-$u$ cuts of $\scri^+$ with the round sphere metric.

In this picture, $\scri^+$ is an affine space, so the choice of origin for the coordinate $u$ is meaningless. This is made manifest by the action of \emph{supertranslations}, which shift $u\to u+f(\lambda,\bar{\lambda})$ for any $f$ valued in $\cO_{\R}(1,1)$; supertranslations shift the origin of $u$ by an arbitrary function of spin weight zero and conformal weight $1$ on the sphere~\cite{Bondi:1962px,Sachs:1962zza}. When $f(\lambda,\bar{\lambda})=a^{\alpha\dot\alpha}\lambda_{\alpha}\bar{\lambda}_{\dot\alpha}$, this is just the action of a Poincar\'e translation.

Of course, this homogeneous formalism applies equally well to advanced Bondi coordinates $(v,r,z,\bar{z})$ and the large-$r$ expansion in the past, for which one obtains a completely analogous description of $\scri^-$ in projective coordinates.

%%%%%%%%%%%%%%%%%%%%%%%%%%

\subsection{Radiative fields and characteristic data at $\scri^+$}
\label{RadField}

A \emph{radiative} field is completely determined by its free characteristic data on $\scri^{\pm}$; such fields are necessarily asymptotically flat vacuum solutions (i.e., source-free) and contain no Coulombic contributions. It is illustrative to consider first the case of a free massless scalar $\Phi$ (of conformal weight -1) in Minkowski space-time; after conformal compactification, an asymptotically flat field admits a large-$r$ expansion
\be\label{scalarex}
\Omega^{-1}\,\Phi(x)=\sum_{k=0}^{\infty}\frac{\varphi^{(k)}(u,z,\bar{z})}{r^{k}}=\sum_{k=0}^{\infty}\frac{\varphi^{(k)}(u,\lambda,\bar{\lambda})}{r^{k}}\,,
\ee
where the coefficient functions $\varphi^{(k)}$ has conformal weight $(-k-1)$ and spin weight zero, or in the homogeneous formalism takes values in $\cO(-k-1,-k-1)$ on $\scri^+$. The field equation can then be considered order-by-order in $r$, so that $\Box\Phi=0$ is translated into a series of evolution equations for the coefficient functions:
\be\label{scalar1}
\dot{\varphi}^{(k+1)}=-\frac{1}{2\,(k+1)}\left(k\,(k+1)+\Delta_{S^2}\right)\varphi^{(k)}\,,
\ee
where $\dot{\varphi}^{(k+1)}:=\partial_{u}\varphi^{(k+1)}$ and $\Delta_{S^2}$ is the Laplacian on the unit sphere.

This means that the leading coefficient function, $\varphi^{(0)}$, is free data on $\scri^+$: all other $\varphi^{(k>0)}$ are determined by $\varphi^{(0)}$ up to $u$-independent functions of integration (cf., \cite{Friedlander:1962lkm,Friedlander:1980}). A \emph{radiative} scalar field is defined by setting all of these functions of integration equal to zero, so that $\Phi$ is completely determined by $\varphi^{(0)}$. In other words, a radiative scalar is one for which $\scri^+$ is a good characteristic data surface, with $\varphi^{(0)}$ the outgoing radiation data. In particular, \emph{any} function $\varphi^{(0)}$ which is smooth and takes values in $\cO(-1,-1)$ (equivalent, has spin weight zero and conformal weight $-1$) defines a radiative solution to the wave equation (at least in a neighbourhood of $\scri^+$)\footnote{The only other condition that one might reasonably require is that the total energy radiated out through $\scri^+$ is finite; this imposes that $\varphi^{(0)}$ fall off at least as quickly as $|u|^{-1}$ as $u\to\pm\infty$}.

\medskip

The notion of radiative field extends to all zero-rest-mass fields of integer or half-integer spin, as well as to \emph{non-linear} gauge and gravitational fields. Consider any asymptotically flat gauge field (with compact gauge group) in Minkowski space; in a radial and asymptotic retarded gauge we have
\[\label{GaugeCondition}
A_r &=& 0  , \\
A_{u} |_{\scri^+} &=& 0.
\]
The gauge potential then admits an asymptotic expansion in retarded homogeneous Bondi coordinates~\cite{vanderBurg:1969,Newman:1978ze,Strominger:2013lka,Barnich:2013sxa}:
\be\label{AFgauge0}
\begin{split}
A_z (r,u,z,\zb) &= A^{(0)}(u,z,\zb)  + \sum_{k=1}^{\infty} \frac{A_z^{(k)}(u,z,\zb)}{r^k}\, ,\\
A_u (r,u,z,\zb) &= \frac{1}{r}\,A_u^{(1)}(u,z,\zb)  + \sum_{k=2}^{\infty} \frac{A_u^{(k)}(u,z,\zb)}{r^{k}}\,.
\end{split}
\ee
From this, it follows that
\be\label{AFgauge}
A|_{\scri^+}=A_z^{(0)}(u,z,\bar{z})\,\d z+A_{\bar{z}}^{(0)}(u,z,\bar{z})\,\d\bar{z}:=\cA^{0}(u,\lambda,\bar{\lambda})\,\D\lambda+\bar{\cA}^{0}(u,\lambda,\bar{\lambda})\,\D\bar{\lambda}\,,
\ee
where $A_z^{(0)}=-\sqrt{2}(1+|z|^2)^{-1}\cA^0$. The reason for introducing the quantity $\cA^0$ will become apparent soon; note that in the homogeneous formalism $\cA^{0}$, $\bar{\cA}^0$ are valued in $\cO(-2,0)$ and $\cO(0,-2)$, respectively, and both take values in the Lie algebra of the gauge group. The six degrees of freedom in the field strength of the gauge field are packaged into the (complex) Newman-Penrose scalars $\Phi_0$, $\Phi_1$ and $\Phi_2$, defined by~\cite{Newman:1961qr}:
\be\label{NPscalars}
\Phi_0:=F_{ab}\,l^{a}\,m^{b}\,, \qquad \Phi_{1}= F_{ab}\left(l^{a}\,n^{b}+m^{a}\,\bar{m}^{b}\right)\,, \qquad \Phi_2=F_{ab}\,\bar{m}^{a}\,n^{b}\,,
\ee
with respect to the Bondi tetrad \eqref{Bonditetrad}. Notice that our definition of the Coulomb scalar $\Phi_1$ differs by a factor of two with respect to the original definition of Newman and Penrose \cite{Newman:1968uj}, such that for a Coulomb charge $q$ we have $\Phi_1=\frac{q}{r^2}+ O(r^{-3})$.

An asymptotically flat gauge field of the form \eqref{AFgauge0} obeys the peeling theorem, meaning that the field strength falls off according to
\[\label{GFpeeling}
\Phi_{0} &=\frac{\phi_0(u,\lambda,\bar{\lambda})}{r^3}+\frac{\phi_0^{(1)}(u,\lambda,\bar{\lambda})}{r^4}+O(r^{-5}) ,\\
\Phi_{1} &= \frac{\phi_1(u,\lambda,\bar{\lambda})}{r^2}+\frac{\phi_1^{(1)}(u,\lambda,\bar{\lambda})}{r^3}+O(r^{-4}),\\
\Phi_{2}&=\frac{\phi_{2}(u,\lambda,\bar{\lambda})}{r}+\frac{\phi_{2}^{(1)}(u,\lambda,\bar{\lambda})}{r^2}+O(r^{-3}).
\]
as $r\to\infty$. Thus, the leading behaviour of the gauge field as one approaches $\scri^+$ (in the conformally-rescaled space-time) is controlled by $\phi_{2}$; for this reason $\phi_2$ is often referred to as the `outgoing radiation field' or `broadcasting function' of the asymptotically flat gauge field. This radiation field is related to the gauge potential itself by
\be\label{Bfunc}
\phi_{2}=\frac{\partial\cA^0}{\partial u}:=\dot{\cA}^0\,,
\ee
so viewed as data on $\scri^+$, $\phi_2$ is a field of spin weight $-1$ and conformal weight $-2$ -- in the homogeneous formalism, this means that $\phi_2$ takes values in $\cO(-3,-1)$. Note that $\phi_2$ is invariant under asymptotic gauge transformations~\cite{Strominger:2013lka}, which shift $\cA^{0}\rightarrow \cA^{0}+\eth\alpha$ for any Lie algebra-valued $\alpha(\lambda,\bar{\lambda})$ taking values in $\cO(0,0)$.

For a general asymptotically flat gauge field, $\phi_2$ gives the radiative parts of the gauge field, while Coulombic contributions are encoded in $\phi_1$ (cf., \cite{Newman:2004ba,Kozameh:2005iw,Adamo:2009vu}). A \emph{radiative gauge field} is one which is completely determined by $\phi_2$ -- such a gauge field will be source-free and consequently have no Coulombic parts. That is, a radiative gauge field is uniquely determined by the characteristic data $\phi_2$ (its outgoing radiation field) on $\scri^+$~\cite{vanderBurg:1969,Exton:1969im,Segal:1979,Ashtekar:1981bq,Friedrich:1986rb}. Specifying any reasonable (i.e., obeying suitable regularity conditions) function valued in $\cO(-3,-1)$ on $\scri^+$ defines a (non-linear) asymptotically flat radiative gauge field. By `reasonable,' we mean that $\phi_2$ is smooth and decays to zero at least as quickly as $|u|^{-3}$ when $u\to\pm\infty$; this latter requirement ensures that the amount of energy carried away by the outgoing radiation field is finite.

\medskip

In the gravitational setting, the Weyl curvature is encoded in the (complex) Newman-Penrose scalars~\cite{Newman:1961qr}
\begin{equation*}
\Psi_{0}:=-C_{abcd}\,l^{a}\,m^{b}\,l^{c}\,m^{d}\,, \qquad \Psi_{1}:=-C_{abcd}\,l^{a}\,n^{b}\,l^{c}\,m^{d}\,,
\end{equation*}
\be\label{NPWeyl}
\Psi_{2}:=-C_{abcd}\,l^{a}\,m^{b}\,\bar{m}^{c}\,n^{d}\,, \qquad \Psi_{3}:=C_{abcd}\,l^{a}\,n^{b}\,n^{c}\,\bar{m}^{d}\,,
\ee
\begin{equation*}
\Psi_{4}:=C_{abcd}\,n^{a}\,\bar{m}^{b}\,n^{c}\,\bar{m}^{d}\,.
\end{equation*}
For any asymptotically flat space-time admitting an expansion \eqref{BS2}, the curvature obeys the peeling theorem:
\[\label{GRpeeling}
\Psi_0 &=\frac{\psi_0(u,\lambda,\bar{\lambda})}{r^5}+\frac{\psi_0^{(1)}(u,\lambda,\bar{\lambda})}{r^6}+O(r^{-7}),\\
\Psi_1 &=\frac{\psi_1(u,\lambda,\bar{\lambda})}{r^4}+\frac{\psi_1^{(1)}(u,\lambda,\bar{\lambda})}{r^5}+O(r^{-6}),\\
\Psi_2 &=\frac{\psi_2(u,\lambda,\bar{\lambda})}{r^3}+\frac{\psi_2^{(1)}(u,\lambda,\bar{\lambda})}{r^4}+O(r^{-5}),\\
\Psi_3 &=\frac{\psi_3(u,\lambda,\bar{\lambda})}{r^2}+\frac{\psi_3^{(1)}(u,\lambda,\bar{\lambda})}{r^3}+O(r^{-4}),\\
\Psi_4 &=\frac{\psi_4(u,\lambda,\bar{\lambda})}{r}+\frac{\psi_4^{(1)}(u,\lambda,\bar{\lambda})}{r^2}+O(r^{-3}).
\]
as $r\to\infty$. So the leading behaviour of the gravitational field near $\scri^+$ (in the conformally re-scaled space-time) is controlled by $\psi_{4}$, which has spin weight $-2$ and conformal weight $-3$. In the homogeneous formalism, this is equivalent to saying that $\psi_4$ takes values in $\cO(-5,-1)$. This gravitational radiation field is related to the data appearing in the asymptotic expansion \eqref{BS2} by
\be\label{Radfunc}
\psi_{4}=-\ddot{\bigma}^{0}:=\dot{N}\,,
\ee
where $N=-\dot{\bigma}^0$ valued in $\cO(-4,0)$ is the news function. Thus, $\psi_4$ is invariant under BMS supertranslations $u\to u+f$ for $f$ in $\cO_{\R}(1,1)$, which act as $\bigma^0\to\bigma^0+\eth^{2}f$~\cite{Sachs:1962zza}.

A purely \emph{radiative} solution of the vacuum Einstein equations is completely characterised by $\psi_4$. Clearly, there are some subtleties arising in the gravitational case, since the existence and smoothness of $\scri^+$ itself depends on the nature of the gravitational field. Friedrich proved that given analytic $\psi_4$, there is a unique solution to the vacuum Einstein equations (at least locally) with this radiation field and a smooth $\scri^+\cong\R\times S^2$~\cite{Friedrich:1986rb}. This proof operates at the level of the characteristic initial value problem for general relativity~\cite{Sachs:1962zzb,Hagen:1977,Friedrich:1983vi}, translated into the language of the conformal Einstein equations (i.e., the vacuum equations lifted to the conformally compactified space-time)~\cite{Friedrich:1981wx,Friedrich:1981at,Kroon:2016ink}. In this context, the characteristic initial value problem can actually be reduced to the standard (and well-posed) Cauchy initial value problem of GR~\cite{Rendall:1990}, with precise existence results~\cite{Luk:2011vf,Chrusciel:2012ap,Friedrich:2013jua,Chrusciel:2013xha,Li:2014mha,Hilditch:2019maz,Zhao:2021mez}.

In other words, given any smooth function valued in $\cO(-5,-1)$ there is a unique radiative space-time in a neighbourhood of $\scri^+$; to ensure that the total gravitational energy radiated out of the space-time is finite, we also impose that $\psi_4$ decay at least as fast as $|u|^{-5}$ as $u\rightarrow\pm\infty$\footnote{The status of this decay condition as $u\to\pm\infty$ is somewhat ambiguous; for different physical motivations one can arrive at quite different bounds (cf., \cite{Christodoulou:1993uv,Campiglia:2015yka,Ashtekar:2018lor}).}. By virtue of being determined entirely by $\psi_4$, radiative space-times are asymptotically flat, source-free and without any Schwarzschild-like (i.e., Coulombic) parts. In addition, the finite energy condition imposes that the Bondi mass aspect $m_{B}\rightarrow 0$ as $u\to\infty$.

\medskip

In the language of the Petrov-Pirani-Penrose classification of algebraically special gauge fields and metrics, it is clear that any type N (i.e., maximally null-degenerate) solution will be a radiative field, since type N gauge fields obey
\be\label{Ngauge}
\Phi_0=\Phi_1=0\,, \qquad \Phi_2\neq 0\,,
\ee
and type N metrics obey
\be\label{Ngrav}
\Psi_0=\Psi_1=\Psi_2=\Psi_3=0\,, \qquad \Psi_4\neq 0\,.
\ee
However, it is important to note that the converse is not true: a radiative field is \emph{not} necessarily algebraically special. Indeed a combination of radial integration of the Newman-Penrose equations and asymptotic Bianchi identities gives the relations for any asymptotically flat Maxwell field in Minkowski space:
\be\label{MBIs}
\dot{\phi}_1=-2\bar{\eth}\phi_2\,, \qquad 2 \dot{\phi}_0=-\bar{\eth}\phi_1\,.
\ee
The radiative condition simply fixes any functions of integration associated with integrating these equations to zero so that everything is determined by $\phi_2$; but clearly this does not force $\phi_1$, $\phi_0$ (and hence $\Phi_1,\Phi_0$) to vanish.

Similarly, radial integrations and asymptotic Bianchi identities for any asymptotically flat metric in Bondi-Sachs form give the relations
\be\label{GRBIs}
\begin{split}
\psi_{3}=-\bar{\eth}N\,, \qquad & \dot{\psi}_2 =-\bar{\eth}\psi_3+\bar{\bigma}^0\,\psi_4\,, \\
\dot{\psi}_{1}=-\bar{\eth}\psi_2+2\,\bar{\bigma}^0\,\psi_3\,, \quad & \dot{\psi}_{0}=-\bar{\eth}\psi_1+3\,\bar{\bigma}^0\,\psi_2\,.
\end{split}
\ee
Once again, the radiative condition sets all functions of integration to zero, but clearly a generic non-vanishing $\psi_4$ will lead to non-vanishing leading coefficient functions for each of the other four Weyl curvature scalars. Even the Bondi mass aspect $m_B$ is generically non-vanishing (for finite $u$), despite the fact that radiative space-times have no Schwarzschild mass contributions. Indeed, the Schwarzschild mass parameter arises as an integration constant (i.e., an $\ell=0$ spherical harmonic contribution to the $\psi_2$ function of integration) when solving the asymptotic Bianchi identities \eqref{GRBIs}; this is set to zero by the radiative condition.

\medskip

%%%

\paragraph{Linear fields in Minkowski space-time:} Clearly, the space of radiative gauge fields and metrics is extremely large: any smooth function of the correct spin and conformal weight (i.e., valued in $\cO(-3,-1)$ or $\cO(-5,-1)$, respectively) with suitable large $|u|$ falloff defines a purely radiative solution. Of course, explicit construction of the non-linear field itself from the characteristic data at $\scri^+$ is not generally possible: given initial data for any hyperbolic non-linear PDE, it will generically be impossible to integrate this data explicitly, even though a solution is guaranteed to be uniquely determined (at least locally). Thus, it is intuitively useful to consider the case of \emph{linearised} radiative fields in Minkowski space, where the field itself can be explicitly constructed directly from the characteristic radiation data.

Any spin $\frac{n}{2}$ (for $n\in\N$) massless free field in four-dimensions can be decomposed into its positive and negative helicity parts (excepting, of course, the $n=0$ case of a massless scalar), represented by the spinor fields $\bar{\phi}_{\dot\alpha_1\cdots\dot\alpha_{n}}$ and $\phi_{\alpha_1\cdots\alpha_n}$, respectively. These are totally symmetric in their spinor indices, and obey the zero-rest-mass equations on-shell:
\be\label{zrm}
\partial^{\alpha\dot\alpha_1}\bar{\phi}_{\dot\alpha_1\cdots\dot\alpha_{n}}=0\,, \qquad \partial^{\alpha_1\dot\alpha}\phi_{\alpha_1\cdots\alpha_n}=0\,.
\ee
As the notation suggests, for Lorentzian-real fields, the positive and negative helicity spinor fields are related by complex conjugation. The characteristic radiative data for such free fields appears in the leading term of their peeling expansion~\cite{Penrose:1965am,Penrose:1986uia}:
\be\label{freepeel}
\begin{split}
\phi_{\alpha_1\cdots\alpha_n}(x)&=\frac{\phi_{n}(u,\lambda,\bar\lambda)}{r}\,\lambda_{\alpha_1}\cdots\lambda_{\alpha_n}+O(r^{-2})\,, \\
\bar{\phi}_{\dot\alpha_1\cdots\dot\alpha_{n}}(x)&=\frac{\bar{\phi}_{n}(u,\lambda,\bar\lambda)}{r}\,\bar{\lambda}_{\dot\alpha_1}\cdots\bar{\lambda}_{\dot\alpha_n}+O(r^{-2})\,,
\end{split}
\ee
with $\phi_n$ taking values in $\cO(-n-1,-1)$. When $n=0$, $\phi_0\equiv\varphi^{(0)}$ from \eqref{scalarex}, when $n=2$ $\phi_2$ is the characteristic data of a radiative Maxwell field from \eqref{Bfunc}, and when $n=4$ $\phi_4\equiv\psi_4$ is the characteristic data of \eqref{Radfunc} now viewed as describing a linear gravitational perturbation of Minkowski space.

To reconstruct the linear field directly from its characteristic data at $\scri^+$, one utilizes a Kirchoff-d'Adh\'emar integral formula adapted to $\scri^+$. In (conformally compactified) Minkowski space-time, every point in the bulk $x\in\M$ is identified with a spherical `cut' $S^2_x\subset\scri^+$, corresponding to where the lightcone with its vertex at $x$ intersects $\scri^+$. In the homogeneous formalism, this cut is described explicitly by
\be\label{lcc}
S^{2}_{x}\,:\quad u=x^{\alpha\dot\alpha}\,\lambda_{\alpha}\,\bar{\lambda}_{\dot\alpha}\,.
\ee
The Kirchoff-d'Adh\'emar integral formula is then given by an integral over such cuts~\cite{Penrose:1980yx,Penrose:1984uia,Penrose:1986uia}:
\be\label{KdA}
\phi_{\alpha_1\cdots\alpha_n}(x)=\int_{S^{2}_x}\D\lambda\wedge\D\bar{\lambda}\,\lambda_{\alpha_1}\cdots\lambda_{\alpha_n}\,\left.\frac{\partial\phi_n}{\partial u}\right|_{S^2_x}\,,
\ee
where the restriction to $S^2_x$ of $\dot\phi_n$ is defined by
\be\label{cutdata}
\left.\frac{\partial\phi_n}{\partial u}(u,\lambda,\bar\lambda)\right|_{S^2_x}:=\frac{\partial\phi_n}{\partial u}(x^{\alpha\dot\alpha}\lambda_\alpha\bar{\lambda}_{\dot\alpha},\lambda,\bar\lambda)\,.
\ee
The version of the formula for the positive helicity part of the field is simply given by complex conjugation of \eqref{KdA}.

It is straightforward to show that fields defined by this integral formula obey the zero-rest-mass equations \eqref{zrm} by differentiating under the integral sign using \eqref{lcc}. The assumption that $\phi_n$ is smooth on $\scri^+$ and decays to zero as $u\to\pm\infty$ ensures that the radiation field of the zero-rest-mass field defined by \eqref{KdA} matches the characteristic data.

%%%%%%%%%%%%%%%%%%%%%%%%%%

\subsection{Double copy prescription}

Consider any radiative gauge field; by definition, this field is uniquely defined by its characteristic data. Making explicit the self-dual and anti-self-dual degrees of freedom, this characteristic data is given by the complex conjugate pair
\be\label{gfradd}
(\phi_{2},\,\bar{\phi}_2)\in \left[\cO(-3,-1)\oplus\cO(-1,-3)\right]\otimes\mathfrak{g}\,,
\ee
where $\mathfrak{g}$ is the Lie algebra of the gauge group. Given any two such radiative gauge fields with the same gauge group, say $(\phi_2^{(1)},\bar{\phi}_{2}^{(1)})$ and $(\phi_2^{(2)},\bar{\phi}_{2}^{(2)})$, we define the \emph{product of characteristic data} as:
\be\label{DCprod}
(\phi_2^{(1)},\,\bar{\phi}_{2}^{(1)})\cdot(\phi_2^{(2)},\,\bar{\phi}_{2}^{(2)}):=
\left(\tr(\phi_2^{(1)}\phi_{2}^{(2)}),\,\tr(\bar{\phi}_{2}^{(1)}\bar{\phi}_{2}^{(2)}),\,\tr(\phi_2^{(1)}\bar{\phi}_{2}^{(2)}),\,\tr(\bar{\phi}_2^{(1)}\phi_{2}^{(2)})\right)\,,
\ee
where the trace is taken over the Lie algebra of the gauge group. In the homogeneous formalism, this product gives a map
\be\label{DCmap}
\left(\left[\cO(-3,-1)\oplus\cO(-1,-3)\right]\otimes\mathfrak{g}\right)^{2}\rightarrow \cO(-6,-2)\oplus\cO(-2,-6)\oplus\cO(-4,-4)\oplus\cO(-4,-4)\,,
\ee
and recalling the correspondence between $\cO(p,q)$ and spin/conformal weight, the individual outputs of the map are valued in
\be\label{DCoutput}
\begin{split}
 \tr(\phi_2^{(1)}\phi_{2}^{(2)})\in\cO(-6,-2)\,, \qquad & s=-2,\; w=-4\,, \\
 \tr(\bar{\phi}_{2}^{(1)}\bar{\phi}_{2}^{(2)})\in\cO(-2,-6)\,, \qquad & s=2\,,\; w=-4\,, \\
 \tr(\phi_2^{(1)}\bar{\phi}_{2}^{(2)}),\,\tr(\bar{\phi}_2^{(1)}\phi_{2}^{(2)})\in\cO(-4,-4)\,, \qquad & s=0\,, \; w=-4\,.
\end{split}
\ee
Thus, the product produces spin weight $\pm2$ functions, as well as two spin weight zero functions on $\scri^+$.

In fact, these outputs canonically define the characteristic data of a radiative metric and two real radiative scalars. First consider $\tr(\phi_2^{(1)}\phi_{2}^{(2)})$ valued in $\cO(-6,-2)$. By the definition of a radiative gauge field, this object is smooth on $\scri^+$ and falls off at least as quickly as $|u|^{-6}$ when $u\to\pm\infty$. An object with the correct spin and conformal weight to be the characteristic data of a radiative metric can therefore be obtained by integrating $\tr(\phi_2^{(1)}\phi_{2}^{(2)})$ once with respect to retarded Bondi time. However, the result is potentially ambiguous, since this integration will produce a function of integration $C_{(-5,-1)}(\lambda,\bar{\lambda})$, where the subscripts denote that this is a function on the sphere valued in $\cO(-5,-1)$.

This ambiguity is fixed by the requirement of finite energy, which forces the gravitational characteristic data to fall off as $|u|^{-5}$ when $u\to\pm\infty$. Meeting this requirement forces us to set $C_{(-5,-1)}=0$, and suitable characteristic data for a radiative metric is obtained by
\be\label{DCmetric}
\psi_{4}(u,\lambda,\bar{\lambda})=\int^{u}\!\d s\,\tr(\phi_2^{(1)}\phi_{2}^{(2)})(s,\lambda,\bar{\lambda})\,,
\ee
with the prescription that the function of integration resulting from the indefinite $u$-integral is set to zero. Applying the same prescription to $\tr(\bar{\phi}_{2}^{(1)}\bar{\phi}_{2}^{(2)})$ gives the complex conjugate $\bar{\psi}_{4}$, as required.

For the scalar degrees of freedom resulting from the product, a similar procedure applies. Characteristic data for a radiative scalar field is obtained by integrating $\tr(\phi_2^{(1)}\bar{\phi}_{2}^{(2)})$ or $\tr(\bar{\phi}_2^{(1)}\phi_{2}^{(2)})$ three times with respect to retarded Bondi time; the resulting function of integration is
\be\label{samb}
C_{(-1,-1)}(\lambda,\bar{\lambda})+u\,C_{(-2,-2)}(\lambda,\bar{\lambda})+u^2\,C_{(-3,-3)}(\lambda,\bar{\lambda})\,,
\ee
with subscripts again denoting scaling weights. This three-fold ambiguity is again fixed by the requirement that the characteristic data falls off as $|u|^{-1}$ when $u\to\pm\infty$, which sets all three functions of integration to zero. With this prescription, the scalar characteristic data is given by
\be\label{DCscalar}
\varphi^{(0)}(u,\lambda,\bar{\lambda})=\iiint^{u} \!\d s\, \tr(\phi_2^{(1)}\bar{\phi}_{2}^{(2)})(s,\lambda,\bar{\lambda})\,,
\ee
and similarly for $\tr(\bar{\phi}_2^{(1)}\phi_{2}^{(2)})$.

\medskip

To summarize, the map \eqref{DCprod} takes \emph{any} two radiative gauge field and produces -- using the prescriptions \eqref{DCmetric}, \eqref{DCscalar} -- a radiative metric and two radiative massless scalar fields. This is precisely as expected for double copy, whereby two gauge fields are combined to yield the NS-NS sector of supergravity: a metric, dilaton and Kalb-Ramond $B$-field. In four-dimensions, the Kalb-Ramond field can be dualized to a scalar axion, so the 4d content of the NS-NS sector is a metric and two scalars.

Let us briefly remark on some basic properties of the map. Firstly, this classical double copy operates entirely at the level of the characteristic data for radiative fields. On the one hand, this is a downside as the fields themselves are never constructed explicitly. On the other hand, this means that the double copy correspondence is \emph{non-linear}, since the characteristic data defines radiative solutions to the fully non-linear equations of motion under consideration.

The map is also essentially independent of the colour structure of the gauge field inputs, which is traced out. In other words, only the functional form of the gauge theory characteristic data is important. Thus, the map is not one-to-one: the same radiative gravitational fields can potentially arise as double copies of various different gauge fields. Neither is this prescription surjective: not every radiative space-time has characteristic data that can be written in the form \eqref{DCmetric}. Yet it should be emphasized that the resulting gravitational fields are not algebraically special, for the same reasons as outlined in Section~\ref{RadField}.

%%%%%%%%%%%%%%%%%%%%%%%%%%
%%%%%%%%%%%%%%%%%%%%%%%%%%

\section{Examples and Relation to Other Classical Double Copies}
\label{RadEx}

As emphasized before, any smooth characteristic data on $\scri^+$ with suitable falloff as $u\to\pm\infty$ defines a radiative scalar, gauge or gravitational field. However, constructing the non-linear fields themselves is virtually impossible, as this entails integrating the non-linear field equations with some generic final data. In this section, we provide several explicit examples of our classical double copy map in cases where the characteristic data can be explicitly integrated. By necessity, such examples are highly non-generic: they are linear, algebraically special or integrable. Finally, we point out a relationship between our double copy prescription restricted to linear fields and another version of classical double copy which was recently formulated in twistor space~\cite{White:2020sfn,Chacon:2021wbr,Chacon:2021hfe}.

%%%%%%%%%%%%%%%%%%%%%%%%%%

\subsection{Momentum eigenstates}

As a first example of our double copy prescription in action, we consider linearised fields in a momentum eigenstate representation. A Maxwell field with null momentum $k^{\alpha\dot\alpha}=\kappa^{\alpha}\,\bar{\kappa}^{\dot\alpha}$ is represented by the zero-rest-mass fields:
\be\label{Maxmeig}
\phi_{\alpha\beta}=\kappa_{\alpha}\,\kappa_{\beta}\,\e^{\im\,k\cdot x}\,, \qquad \bar{\phi}_{\dot\alpha\dot\beta}=\bar{\kappa}_{\dot\alpha}\,\bar{\kappa}_{\dot\beta}\,\e^{\im\,k\cdot x}\,,
\ee
which solve the linear free field equations \eqref{zrm} for $n=2$. The characteristic data associated to such a momentum eigenstate is
\be\label{Maxdata}
\phi_{2}=-\im\,\frac{\la a\,\kappa\ra^2}{\la a\,\lambda\ra^2}\,\delta\!\left(\mathrm{Re}\,\la\lambda\,\kappa\ra\right)\,\delta\!\left(\mathrm{Im}\,\la\lambda\,\kappa\ra\right)\,\exp\left(\im\,u\,\frac{\la a\,\kappa\ra\,[\bar{a}\,\bar{\kappa}]}{\la a\,\lambda\ra\,[\bar{a}\,\bar{\lambda}]}\right)\,,
\ee
which is valued in $\cO(-3,-1)$ as required. Here, $a^{\alpha}$, $\bar{a}^{\dot\alpha}$ are arbitrary constant spinors; as such they do not appear in the gauge-invariant zero-rest-mass fields \eqref{Maxmeig}.

To see that this is correct, it suffices to show that \eqref{Maxmeig} is reproduced when $\phi_2$ is inserted into the Kirchoff-d'Adh\'emar integral formula \eqref{KdA}. For $\phi_2$ given by \eqref{Maxdata}, observe that
\be\label{mKdA1}
\D\bar{\lambda}\,\dot{\phi}_2=\frac{\la a\,\kappa\ra^3\,[\bar{a}\,\bar{\kappa}]}{\la a\,\lambda\ra^3\,[\bar{a}\,\bar{\lambda}]}\,\D\bar{\lambda}\,\delta\!\left(\mathrm{Re}\,\la\lambda\,\kappa\ra\right)\,\delta\!\left(\mathrm{Im}\,\la\lambda\,\kappa\ra\right)\,\exp\left(\im\,u\,\frac{\la a\,\kappa\ra\,[\bar{a}\,\bar{\kappa}]}{\la a\,\lambda\ra\,[\bar{a}\,\bar{\lambda}]}\right)\,.
\ee
Now, the Schouten identity gives
\be\label{Schouten}
\D\bar{\lambda}\,[\bar{a}\,\bar{\kappa}]=[\bar{\lambda}\,\d\bar{\lambda}]\,[\bar{a}\,\bar{\kappa}]=[\bar{\lambda}\,\bar{\kappa}]\,[\bar{a}\,\d\bar{\lambda}]+[\bar{a}\,\bar{\lambda}]\,[\bar{\kappa}\,\d\bar{\lambda}]\,,
\ee
which can be substituted into \eqref{mKdA1}. On the support of the delta functions, the first term on the RHS of this Schouten identity vanishes, leaving
\be\label{mKdA2}
\begin{split}
\D\bar{\lambda}\,\dot{\phi}_2 &=\frac{\la a\,\kappa\ra^3\,[\bar{\kappa}\,\d\bar{\lambda}]}{\la a\,\lambda\ra^3}\,\delta\!\left(\mathrm{Re}\,\la\lambda\,\kappa\ra\right)\,\delta\!\left(\mathrm{Im}\,\la\lambda\,\kappa\ra\right)\,\exp\left(\im\,u\,\frac{\la a\,\kappa\ra\,[\bar{a}\,\bar{\kappa}]}{\la a\,\lambda\ra\,[\bar{a}\,\bar{\lambda}]}\right) \\
 &=\frac{1}{2\pi\im}\,\frac{\la a\,\kappa\ra^3}{\la a\,\lambda\ra^3}\,\d\bar{\lambda}^{\dot\alpha}\,\frac{\partial}{\partial\bar{\lambda}^{\dot\alpha}}\left(\frac{1}{\la\lambda\,\kappa\ra}\right)\,\exp\left(\im\,u\,\frac{\la a\,\kappa\ra\,[\bar{a}\,\bar{\kappa}]}{\la a\,\lambda\ra\,[\bar{a}\,\bar{\lambda}]}\right) \\
  &= \frac{\la a\,\kappa\ra^3}{\la a\,\lambda\ra^3}\,\bar{\delta}\!\left(\la\lambda\,\kappa\ra\right)\,\exp\left(\im\,u\,\frac{\la a\,\kappa\ra\,[\bar{a}\,\bar{\kappa}]}{\la a\,\lambda\ra\,[\bar{a}\,\bar{\lambda}]}\right)\,,
\end{split}
\ee
where we have used the definition of the holomorphic delta function $\bar{\delta}(z):= (2\pi\im)^{-1} \dbar(z^{-1})=\d\bar{z}\,\delta(\mathrm{Re}\,z)\,\delta(\mathrm{Im}\,z)$. Feeding this into the Kirchoff-d'Adh\'emar integral formula gives
\be\label{mKdA3}
\begin{split}
\phi_{\alpha\beta} & =\int_{S^2_x}\D\lambda\wedge\D\bar{\lambda}\,\lambda_{\alpha}\,\lambda_{\beta}\,\dot{\phi}_2|_{S^2_x} \\
 & = \int_{S^2_x}\lambda_{\alpha}\,\lambda_{\beta}\, \frac{\la a\,\kappa\ra^3}{\la a\,\lambda\ra^3}\,\D\lambda\wedge\bar{\delta}\!\left(\la\lambda\,\kappa\ra\right)\,\exp\left(\im\,\la\lambda|x|\bar{\lambda}]\,\frac{\la a\,\kappa\ra\,[\bar{a}\,\bar{\kappa}]}{\la a\,\lambda\ra\,[\bar{a}\,\bar{\lambda}]}\right) \\
  & = \kappa_{\alpha}\,\kappa_{\beta}\,\e^{\im\,k\cdot x}\,,
\end{split}
\ee
as required. A similar calculation works for the positive helicity part of the Maxwell field.

Although we have successfully constructed the characteristic data for the momentum eigenstate, this is not actually `good' data for an asymptotically flat linear field. Indeed, $\phi_2$ is certainly not smooth, having delta-function singularities on the generator of $\scri^+$ corresponding to the null momentum, and is also oscillatory in $u$ rather than decaying like $|u|^{-3}$ as $u\to\pm\infty$. The latter problem is easily resolved with the usual $\im\varepsilon$-prescription; namely, for asymptotic values of $u$ we should deform $\im\,u\rightarrow (\im -\varepsilon)\,u$ for some small $\varepsilon>0$, so that the data is exponentially suppressed.

The singularity on the celestial sphere at $(\lambda,\bar{\lambda})=(\kappa,\bar{\kappa})$ is more difficult to deal with. We can regularize the delta functions in \eqref{Maxdata} by making them finite with compact support, for instance with rectangular functions:
\be\label{deltareg}
\delta(x)\rightarrow\delta_{\varepsilon}(x):=\left\{\begin{array}{r}
																		\frac{1}{\varepsilon} \quad \mbox{if } x\in(-\frac{\varepsilon}{2},\frac{\varepsilon}{2}) \\
																		0 \quad \mbox{otherwise}
																	\end{array}\right.\,,
\ee
although any other `thickened' representation of a delta function will suffice. The main property we require of the regularisation is that $\delta_{\varepsilon}(x)\,\delta_{\varepsilon}(x)=f(x)\,\delta_{2\varepsilon}(x)$ where $f(x)$ is any even function which obeys $f(ax)=|a|^{-1}f(x)$ and $f(0)=1$. This ensures that the regularised product of two delta functions has the correct scaling and symmetry properties. Since the limit $\varepsilon\to0$ does not commute with multiplication, this allows us to obtain a delta function from a product of two regularised delta functions, in contrast to the product of two (unregulated) delta functions, which is zero as a distribution.

\medskip

This regularisation prescription is only sufficient to allow the double copy map to act on symmetric inputs: that is, two copies of the \emph{same} gauge theoretic momentum eigenstate. If we chose two different momentum eigenstates, the inputs are singular on distinct generators of $\scri^+$ and the regularisation mechanism is not sufficient to produce sensible outputs. Thus, we are left to consider $(\phi_2,\bar{\phi}_2)\cdot(\phi_2,\bar{\phi}_2)$; the first output:
\be\label{medc1}
\phi_2^2=-\lim_{\varepsilon\to 0}\frac{\la a\,\kappa\ra^5\,[\bar{a}\,\bar{\kappa}]}{\la a\,\lambda\ra^5\,[\bar{a}\,\bar{\lambda}]}\,\delta_{2\varepsilon}\!\left(\mathrm{Re}\,\la\lambda\,\kappa\ra\right) \delta_{2\varepsilon}\!\left(\mathrm{Im}\,\la\lambda\,\kappa\ra\right)\,\exp\left(2\im\,u\,\frac{\la a\,\kappa\ra\,[\bar{a}\,\bar{\kappa}]}{\la a\,\lambda\ra\,[\bar{a}\,\bar{\lambda}]}\right)\,,
\ee
is valued in $\cO(-6,-2)$. Here, we used our regulation procedure to set
\be\label{deltareg2}
\delta^2_{\varepsilon}\!\left(\mathrm{Re}\,\la\lambda\,\kappa\ra\right) \delta^{2}_{\varepsilon}\!\left(\mathrm{Im}\,\la\lambda\,\kappa\ra\right)=\frac{[\bar{a}\,\bar{\kappa}]}{[\bar{a}\,\bar{\lambda}]}\,\delta_{2\varepsilon}\!\left(\mathrm{Re}\,\la\lambda\,\kappa\ra\right) \delta_{2\varepsilon}\!\left(\mathrm{Im}\,\la\lambda\,\kappa\ra\right)\,.
\ee
Upon taking $\varepsilon\to0$ and using the double copy \eqref{DCmetric}, one obtains
\be\label{megrav1}
\psi_{4}=\frac{\im}{2}\,\frac{\la a\,\kappa\ra^4}{\la a\,\lambda\ra^4}\,\delta\!\left(\mathrm{Re}\,\la\lambda\,\kappa\ra\right) \delta\!\left(\mathrm{Im}\,\la\lambda\,\kappa\ra\right)\,\exp\left(2\im\,u\,\frac{\la a\,\kappa\ra\,[\bar{a}\,\bar{\kappa}]}{\la a\,\lambda\ra\,[\bar{a}\,\bar{\lambda}]}\right)\,.
\ee
Feeding this into the Kirchoff-d'Adh\'emar integral formula (with $n=4$) gives the linearised space-time field
\be\label{megrav2}
\phi_{\alpha\beta\gamma\delta}=\int_{S^2_x}\D\lambda\wedge\D\bar{\lambda}\,\lambda_{\alpha}\,\lambda_{\beta}\,\lambda_{\gamma}\,\lambda_{\delta}\,\dot{\psi}_{4}|_{S^2_x}=\frac{\im}{2}\,\kappa_{\alpha}\,\kappa_{\beta}\,\kappa_{\gamma}\,\kappa_{\delta}\,\e^{2\im\,k\cdot x}\,,
\ee
which is a negative helicity gravitational momentum eigenstate. Observe that the momentum of the gravitational state has picked up a factor of 2 relative to the initial gauge theory input; in other double copy constructions, this would be normalised by division against a scalar mode but this is not necessary to produce a solution to the gravitational zero-rest-mass equations.

The computation of $\bar{\phi}^{2}_{2}$ proceeds along similar lines, producing the positive helicity counterpart of \eqref{megrav2}. The double copy also gives two copies of the spin-weight zero data
\be\label{mescalar1}
|\phi_2|^2=\frac{\la a\,\kappa\ra^3\,[\bar{a}\,\bar{\kappa}]^{3}}{\la a\,\lambda\ra^{3}\,[\bar{a}\,\bar{\lambda}]^3}\,\delta\!\left(\mathrm{Re}\,\la\lambda\,\kappa\ra\right)\, \delta\!\left(\mathrm{Im}\,\la\lambda\,\kappa\ra\right)\,\exp\left(2\im\,u\,\frac{\la a\,\kappa\ra\,[\bar{a}\,\bar{\kappa}]}{\la a\,\lambda\ra\,[\bar{a}\,\bar{\lambda}]}\right)\,,
\ee
valued in $\cO(-4,-4)$, after using the same regularisation procedure as before. The prescription \eqref{DCscalar} then produces the characteristic radiative data
\be\label{mescalar2}
\varphi^{(0)}=\frac{\im}{8}\,\delta\!\left(\mathrm{Re}\,\la\lambda\,\kappa\ra\right)\, \delta\!\left(\mathrm{Im}\,\la\lambda\,\kappa\ra\right)\,\exp\left(2\im\,u\,\frac{\la a\,\kappa\ra\,[\bar{a}\,\bar{\kappa}]}{\la a\,\lambda\ra\,[\bar{a}\,\bar{\lambda}]}\right)\,.
\ee
Feeding this into the Kirchoff-d'Adh\'emar integral formula (with $n=0$) gives
\be\label{mescalar3}
\phi=\int_{S^2_x}\D\lambda\wedge\D\bar{\lambda}\,\dot{\varphi}^{(0)}|_{S^2_x}=\frac{\im}{8}\,\e^{2\im\,k\cdot x}\,,
\ee
which is the expected massless scalar momentum eigenstate. Once again, note that the momentum has picked up a factor of 2 relative to the gauge theory momentum eigenstates. This indicates that the notion of double copy defined by the map \eqref{DCprod} is not directly related to the usual one arising from scattering amplitudes, where this doubling of momentum does not occur.

%%%%%%%%%%%%%%%%%%%%%%%%%%

\subsection{Plane waves}

Our next example is provided by \emph{plane wave} solutions of Yang-Mills theory, which admit a covariantly constant null symmetry generated by a null vector $n$ and have a 5-dimensional Heisenberg algebra of symmetries whose centre is $n$. These symmetry constraints actually force the solution to take values in the Cartan subalgebra of the gauge group~\cite{Trautman:1980bj,Adamo:2017nia}. With the Minkowski metric written in light-front coordinates
\be\label{lfMink}
\d s^2=2\left(\d x^+\,\d x^- - \d z\,\d\bar{z}\right)\,,
\ee
a plane wave gauge potential can be written as:
\be\label{PWgauge}
A=\left(f'(x^-)\,z+\bar{f}'(x^-)\,\bar{z}\right)\d x^-\,,
\ee
where the profile function $f(x^-)$ is an arbitrary complex function of lightfront time valued in the Cartan of the gauge group and $f':=\partial_{-}f$. The covariantly constant null symmetry associated with this gauge potential is generated by $n=\partial_{+}$, which we can represent in 2-spinors by choosing a spinor dyad $(o^{\alpha},\iota^{\alpha})$ (normalised so that $\la\iota\,o\ra=1$) and its complex conjugate such that $n^{\alpha\dot\alpha}=\iota^{\alpha}\,\bar{\iota}^{\dot\alpha}$. The self-dual and anti-self-dual parts of the field strength associated to \eqref{PWgauge} are then
\be\label{PWfs}
\bar{\phi}_{\dot\alpha\dot\beta}=\bar{\iota}_{\dot\alpha}\,\bar{\iota}_{\dot\beta}\,\bar{f}'\,, \qquad \phi_{\alpha\beta}=\iota_{\alpha}\,\iota_{\beta}\,f'\,,
\ee
demonstrating that such plane waves are type N: that is, maximally null degenerate and algebraically special. It is easy to see that such a plane wave is a solution to the vacuum Yang-Mills equations for any choice of $f(x^-)$ valued in the Cartan.

By virtue of being Cartan-valued, the Kirchoff-d'Adh\'emar integral formula \eqref{KdA} can be used to determine the characteristic data for the plane wave. In particular, one finds that
\be\label{PWgdata}
\phi_2=\frac{1}{\la\lambda\,o\ra^3\,[\bar{\lambda}\,\bar{o}]}\,\delta\!\left(\mathrm{Re}\,\frac{\la\iota\,\lambda\ra}{\la o\,\lambda\ra}\right)\,\delta\!\left(\mathrm{Im}\,\frac{\la\iota\,\lambda\ra}{\la o\,\lambda\ra}\right)\,f\!\left(\frac{u}{\la\lambda\,o\ra\,[\bar{\lambda}\,\bar{o}]}\right)\,,
\ee
taking values in $\cO(-3,-1)$ and the Cartan subalgebra. It is straightforward to verify that this $\phi_2$ produces \eqref{PWfs} by making use of the Schouten identity
\be\label{Schouten2}
\D\bar{\lambda}=\D\bar{\lambda}\,[\bar{\iota}\,\bar{o}]=[\bar{\lambda}\,\bar{\iota}]\,[\d\bar{\lambda}\,\bar{o}]+[\bar{\lambda}\,\bar{o}]\,[\bar{\iota}\,\d\bar{\lambda}]\,,
\ee
as well as the support of the delta functions in \eqref{PWgdata} and the definition of the holomorphic delta function to find
\be\label{PWgdata2}
\D\bar{\lambda}\,\dot{\phi}_2=\frac{1}{\la\lambda\,o\ra^4}\,\bar{\delta}\!\left(\frac{\la\lambda\,\iota\ra}{\la\lambda\,o\ra}\right)\,f^{\prime}\!\left(\frac{u}{\la\lambda\,o\ra\,[\bar{\lambda}\,\bar{o}]}\right)\,,
\ee
which can be fed directly into the Kirchoff-d'Adh\'emar integral.

Now, it is easy to see that the same problems which arose with momentum eigenstates arise for the plane wave: the field is not actually asymptotically flat! To be able to treat these fields with our double copy prescription, we follow a similar strategy to that applied to momentum eigenstates. First, we restrict our attention to `sandwich' plane waves for which $f(x^-)$ is smooth and compactly supported in lightfront time\footnote{This is slightly stronger than the usual sandwich condition, which is that the field strength of the plane wave is compactly supported in lightfront time (cf., ~\cite{Bondi:1958aj}). This weaker condition only imposes that $f'(x^-)$ be compactly supported, whereas we are requiring that $\phi_2$ -- and hence $f(x^-)$ itself -- has compact support.}. This ensures that the plane wave carries a finite amount of energy.

But even with this condition, the characteristic data is still singular along the null generator $(\lambda,\bar{\lambda})=(\iota,\bar{\iota})$ of $\scri^+$ corresponding to the null symmetry of the plane wave. This issue can be dealt with using the same `regularisation' procedure that we applied to the momentum eigenstates. This means that we can only take the double copy of plane waves with the \emph{same} null symmetry $n^{\alpha\dot\alpha}=\iota^{\alpha}\bar{\iota}^{\dot\alpha}$, although the compactly supported profile functions can be different.

With this in mind, consider two plane wave gauge fields with sandwich profile functions valued in the Cartan subalgebra of some compact gauge group (the same for both fields); denote these profile functions by $f$ and $g$ and their characteristic data by $\phi_2^f$, $\phi_2^g$, respectively. Then the component of the double copy map which encodes spin $-2$ data yields
\be\label{PWgrav1}
\tr\left(\phi_2^f\,\phi_2^g\right)=\frac{1}{\la\lambda\,o\ra^{6}\,[\bar{\lambda}\,\bar{o}]^2} \,\delta\!\left(\mathrm{Re}\,\frac{\la\iota\,\lambda\ra}{\la o\,\lambda\ra}\right)\,\delta\!\left(\mathrm{Im}\,\frac{\la\iota\,\lambda\ra}{\la o\,\lambda\ra}\right)\,\mathscr{F}\!\left(\frac{u}{\la\lambda\,o\ra\,[\bar{\lambda}\,\bar{o}]}\right)\,,
\ee
where
\be
\mathscr{F}(x^-):=\tr\left[f(x^-)\, g(x^-)\right]\,,
\ee
with the trace running over the Cartan. Using the prescription \eqref{DCmetric} gives the gravitational characteristic data:
\be\label{PWgrav2}
\psi_{4}= \frac{1}{\la\lambda\,o\ra^{5}\,[\bar{\lambda}\,\bar{o}]} \,\delta\!\left(\mathrm{Re}\,\frac{\la\iota\,\lambda\ra}{\la o\,\lambda\ra}\right)\,\delta\!\left(\mathrm{Im}\,\frac{\la\iota\,\lambda\ra}{\la o\,\lambda\ra}\right)\,\int^{\frac{u}{\la\lambda\,o\ra\,[\bar{\lambda}\,\bar{o}]}}\mathscr{F}(s)\,\d s\,.
\ee
The anti-self-dual gravitational data $\bar{\psi}_4$ is easily seen to be the complex conjugate of this expression.

Since both $f$ and $g$ are compactly supported, this ensures that the gravitational data is also compactly supported. The data $\psi_4$ is not smooth over all of $\scri^+$, but its singularity on the null generator $(\lambda,\bar{\lambda})=(\iota,\bar{\iota})$ is of precisely the same form as the initial gauge theoretic plane wave data \eqref{PWgdata}. In particular, it is easy to see that \eqref{PWgrav2} is in fact the characteristic data for a plane wave \emph{metric}, with line element
\be\label{PWgravmet}
\d s^2=2\left(\d x^{+}\,\d x^{-}-\d z\,\d\bar{z}\right)-\left(z^2\,\tr\left[f(x^-)\, g(x^-)\right]+\bar{z}^2\,\tr\left[\bar{f}(x^-)\,\bar{g}(x^-)\right]\right)\,(\d x^-)^2\,,
\ee
in Brinkmann coordinates. This is a solution of the vacuum Einstein equations for any $f,g$ with Weyl tensor given by
\be\label{PWWeyl}
\Psi_{\alpha\beta\gamma\delta}=\iota_{\alpha}\,\iota_{\beta}\,\iota_{\gamma}\,\iota_{\delta}\,\tr\left[f(x^-)\, g(x^-)\right]\,,
\ee
and is thus algebraically special of type N.

Following a similar procedure for the scalar degrees of freedom produced by \eqref{DCprod} -- \eqref{DCmap} and using the prescription \eqref{DCscalar} gives the scalar characteristic data:
\be\label{PWscalar1}
\varphi^{(0)}= \frac{1}{\la\lambda\,o\ra\,[\bar{\lambda}\,\bar{o}]} \,\delta\!\left(\mathrm{Re}\,\frac{\la\iota\,\lambda\ra}{\la o\,\lambda\ra}\right)\,\delta\!\left(\mathrm{Im}\,\frac{\la\iota\,\lambda\ra}{\la o\,\lambda\ra}\right)\,\iiint^{\frac{u}{\la\lambda\,o\ra\,[\bar{\lambda}\,\bar{o}]}}\tr\left[f(s)\,\bar{g}(s)\right]\,\d s\,.
\ee
As expected, this corresponds to the complex scalar
\be\label{PWscalar2}
\Phi(x)=\iint^{x^{-}}\tr\left[f(s)\,\bar{g}(s)\right]\,\d s\,,
\ee
which solves the massless wave equation for any $f,g$.

Thus, our prescription states that the classical double copy of a gauge-theoretic plane wave is a NS-NS plane wave: namely, a gravitational plane wave and two (real) scalar plane waves. This is the same correspondence that emerges with the Kerr-Schild classical double copy~\cite{Monteiro:2014cda,Godazgar:2020zbv}, albeit in a very different fashion.

%%%%%%%%%%%%%%%%%%%%%%%%%%

\subsection{Self-dual radiative fields}

In Lorentzian signature, the self-dual and anti-self-dual parts of gauge and gravitational field are related by complex conjugation. However, for complex fields (or real-valued fields in Euclidean or ultrahyperbolic signature) these become distinct, independent degrees of freedom, allowing for non-trivial purely \emph{self-dual} configurations. A self-dual radiative gauge field (cf., \cite{Goldberg:1979wt,Newman:1980fr,Adamo:2020yzi}) is a complex radiative gauge field whose characteristic data at $\scri^+$ takes the form $\phi_2=0$, $\tilde{\phi_2}\neq0$, where we use the notation $\tilde{\phi_2}$ to emphasize that this is no longer the complex conjugate of $\phi_2$.

For such self-dual radiative gauge fields, the double copy map \eqref{DCprod} is significantly simplified:
\be\label{SDmap}
(0,\,\tilde{\phi}^{(1)}_{2})\cdot(0,\,\tilde{\phi}^{(2)}_2)=\left(0\,,\tr(\tilde{\phi}^{(1)}_{2}\tilde{\phi}^{(2)}_{2}),\,0,\,0\right)\,,
\ee
and thus only produces gravitational data $\tilde{\psi}_{4}$. A complex radiative space-time whose characteristic data at $\scri^+_\C$ (the complexification of $\scri^+$ by allowing for $u\in\C$) is $\psi_4=0$, $\tilde{\psi}_{4}\neq0$ is a self-dual radiative space-time, also refered to in the literature as a $\cH$-space (cf., \cite{Newman:1976gc,Ko:1981}). Such complex radiative space-times can still be characterised as asymptotically flat, with appropriate definitions~\cite{Ludvigsen:1981}. Hence, our classical double copy takes any self-dual radiative gauge field and produces a self-dual radiative metric, with the two scalar degrees of freedom being trivial.

While non-linear, these solutions can be explictly reconstructed from their characteristic data using twistor theory to exploit the underlying integrability of the self-dual sector~\cite{Ward:1977ta,Penrose:1976js,Mason:1991rf}. For a self-dual radiative gauge field, the bulk gauge field is constructed directly from $\tilde{\cA}^0$ (i.e., the first $u$-integral of $\tilde{\phi}_2$) by means of the Sparling equation~\cite{Newman:1980fr,Sparling:1990}:
\be\label{Sparlingeq}
\frac{\partial H}{\partial\bar{\lambda}^{\dot\alpha}}(x,\lambda,\bar{\lambda})=-\bar{\lambda}_{\dot\alpha}\,\tilde{\cA}^{0}|_{S^2_x}\,H(x,\lambda,\bar{\lambda})\,,
\ee
which is a PDE on the sphere of generators of $\scri^+$ for a frame $H$ valued in the Lie algebra of the gauge group. The space-time gauge potential is implicitly encoded in this $H$ via
\be\label{Frame}
H^{-1}\,\lambda^{\alpha}\partial_{\alpha\dot\alpha}H=\im\,\lambda^{\alpha}\,A_{\alpha\dot\alpha}\,.
\ee
In general, it can be difficult to solve for $H$ -- and hence $A_{\alpha\dot\alpha}$ -- explicitly for generic self-dual characteristic data, but for some particular examples this is possible.

Consider the abelian data:
\be\label{SDab1}
\tilde{\cA}^{0}=\frac{a}{u^2\,(1+|z|^2)}=a\,\frac{\la\lambda\,o\ra^2}{u^2}\,,
\ee
where we use a spinor dyad $o_{\alpha}=(1,0)$, $\iota_{\alpha}=(0,1)$ to express the data in homogeneous coordinates and $a$ is an overall constant parametrising the field strength, into which we absorb any numerical factors. For this data, the abelian Sparling equation is easily solved with
\be\label{SDab2}
H(x,\lambda,\bar\lambda)=\exp[-g(x,\lambda,\bar{\lambda})]\,, \qquad g=\frac{2\,a}{x^2}\,\frac{\la\lambda\,o\ra\,\la o|x|\bar{\lambda}]}{\la\lambda|x|\bar{\lambda}]}\,,
\ee
from which the gauge potential is recovered from \eqref{Frame}
\be\label{SDab3}
A_{\alpha\dot\alpha}(x)=-2\im\,a\,\frac{x^{\beta}{}_{\dot\alpha}\,o_{\alpha}\,o_{\beta}}{x^4}\,.
\ee
From this, one easily confirms that the field is self-dual
\be\label{SDab4}
\phi_{\alpha\beta}=0\,, \quad \tilde{\phi}_{\dot\alpha\dot\beta}=\frac{\im\,a}{x^2}\,\tilde{s}_{\dot\alpha}\,\tilde{s}_{\dot\beta}\,, \qquad \mbox{for } \quad \tilde{s}_{\dot\alpha}:=2\,\frac{x^{\beta\dot\alpha}\,o_{\beta}}{x^2}\,,
\ee
and therefore maximally null-degenerate.

The characteristic radiative data associated with \eqref{SDab1} is
\be\label{SDabdata}
\tilde{\phi}_2=-2\,a\,\frac{\la\lambda\,o\ra^2}{u^3}\,,
\ee
which is regular on the sphere and has the required falloff as $u\to\pm\infty$. However, there is also clearly a singularity at $u=0$ -- this is related to the fact that the gauge potential \eqref{SDab3} is singular on the lightcone $x^2=0$. Since $u$ is merely an affine coordinate, we are free to shift it by a (complex) Poincar\'e translation $u\to u +a^{\alpha\dot\alpha}\lambda_{\alpha}\bar{\lambda}_{\dot\alpha}$, thereby removing the singularity. We will not do this explictly, instead keeping in mind that this $u=0$ singularity is spurious.

With this in mind, consider two such self-dual radiative Maxwell fields, with potentially distinct overall constants $a,b$. The double copy then produces self-dual gravitational characteristic data
\be\label{SDgrav1}
\tilde{\psi}_{4}=-\frac{a\,b}{5}\,\frac{\la\lambda\,o\ra^4}{u^5}\,.
\ee
This is `good' radiative data, in the sense that it is regular on the sphere, has the required fall-off as $u\to\pm\infty$; the spurious singularity at $u=0$ is only an artifact of the analogous singularity in the gauge theory data. The self-dual radiative metric corresponding to this data can be recovered by integrating twice to obtain the asymptotic shear
\be\label{SDgrav2}
\tilde{\bigma}^0=\frac{c}{u^3\,(1+|z|^2)^2}=\frac{c\,\la\lambda\,o\ra^4}{u^3}\,,
\ee
where $c:=-\frac{ab}{60}$, and feeding this into the good cut equation~\cite{Newman:1976gc,Eastwood:1982}:
\be\label{GCE}
\frac{\partial^2 Z(\lambda,\bar{\lambda})}{\partial\bar{\lambda}^{\dot\alpha}\,\partial\bar{\lambda}^{\dot\beta}}=\bar{\lambda}_{\dot\alpha}\,\bar{\lambda}_{\dot\beta}\,\tilde{\bigma}^{0}(Z,\lambda,\bar{\lambda})\,.
\ee
This equation determines spherical cross sections of $\scri^+_{\C}$ corresponding to asymptotically shear free null geodesic congruences in terms of the cut function $u=Z(\lambda,\bar{\lambda})$, from which a metric can be derived.

As a non-linear PDE, the good cut equation is usually impossible to solve (although a four-parameter family of solutions is always guaranteed to exist for sufficiently `small' data $\tilde{\bigma}^0$~\cite{Hansen:1978jz}). Fortunately, the good cut equation with data given by \eqref{SDgrav2} has already been solved in the literature~\cite{Sparling:1981nk}:
\be\label{ST}
Z(x,\lambda,\bar{\lambda})=\la\lambda|x|\bar{\lambda}]+c\,\la\lambda\,o\ra\,[\bar{\lambda}\,\tilde{s}]\,,
\ee
with $\tilde{s}^{\dot\alpha}$ defined by \eqref{SDab4}. The metric obtained from this cut function, written in a null coordinate system $(u,X,Y,v)$, is Kerr-Schild~\cite{Sparling:1981nk}:
\be\label{EH}
\d s^2=2\left(\d u\,\d v-\d X\,\d Y\right)-\frac{2\,c}{(u\,v-X\,Y)^3}\left(Y\,\d v-v\,\d Y\right)^2\,,
\ee
and is easily verified to be a self-dual type N vacuum solution~\cite{Tod:1982mmp}. This metric can be viewed as a type N analogue of the Eguchi-Hanson ALE gravitational instanton (which is type D) when written in Kerr-Schild coordinates (cf., \cite{Hitchin:1900zr,Tod:1982mmp}). The Eguchi-Hanson metric can also be obtained using Kerr-Schild~\cite{Berman:2018hwd} and Weyl~\cite{Luna:2018dpt} double copies; in the latter there are two ways of doing this from a `symmetric' or `mixed' gauge field. While the gauge field \eqref{SDab3} is different from that appearing in~\cite{Berman:2018hwd} or the `symmetric' approach of~\cite{Luna:2018dpt}, it agrees with one of the `mixed' single copy gauge fields found using the Weyl double copy~\cite{Luna:2018dpt}!

\medskip

Another well-known example of a self-dual radiative gauge field for which the Sparling equation \eqref{Sparlingeq} can be solved explicitly is the SU$(2)$ instanton; the self-dual field strength of the non-abelian gauge field is
\be\label{inst1}
\tilde{\Phi}_{\dot\alpha\dot\beta}=\frac{4\,a\,\delta^{(\dot\gamma}_{\dot\alpha}\,\delta^{\dot\delta)}_{\dot\beta}}{(1+a\,x^2)^2}\,,
\ee
where $a$ is the instanton size parameter and the symmetric pair of indices $(\dot\gamma\dot\delta)$ are identified with a SU$(2)$ Lie algebra index. Using the Sparling equation, or by analyzing the large $r$ behaviour of the field strength, one obtains the characteristic radiative data~\cite{Hsieh:1986aq}
\be\label{inst2}
\tilde{\cA}^{0}=\frac{t_{\alpha}{}^{(\dot\gamma}\,t_{\beta}{}^{\dot\delta)}\,\lambda^{\alpha}\,\lambda^{\beta}}{a\,u^2}\,, \qquad \tilde{\phi}_2=-2\,\frac{t_{\alpha}{}^{(\dot\gamma}\,t_{\beta}{}^{\dot\delta)}\,\lambda^{\alpha}\,\lambda^{\beta}}{a\,u^3}\,.
\ee
This is `good' radiative data (smooth on the sphere and with appropriate asymptotic $u$ behaviour) apart from the singularity at $u=0$; much like our previous examples, this singularity is spurious and can be removed (cf., \cite{Jiang:2008xw}). Unfortunately, this non-abelian radiative data is null with respect to the double copy map, since
\be\label{inst3}
\tr(\tilde{\phi}_2^2)=\frac{\epsilon_{\alpha\delta}\,\epsilon_{\beta\gamma}}{a^2\,u^6}\,\lambda^{\alpha}\,\lambda^{\beta}\,\lambda^{\gamma}\,\lambda^{\delta}=0\,.
\ee
Thus, this simplest explicit non-abelian example of a self-dual radiative gauge field fails to produce a non-trivial self-dual radiative gravitational field with this double copy prescription.

%%%%%%%%%%%%%%%%%%%%%%%%%%

\subsection{Spherical waves}

As a final example, consider a spherical, purely radiative wave in Maxwell theory; the field strength is expressed in the Newman-Penrose formalism and standard (affine) Bondi coordinates $(u,r,z,\bar{z})$ by (cf., Chapter 28.2 of \cite{Stephani:2003}):
\be\label{swave1}
\Phi_0=0\,, \qquad \Phi_1=\frac{Q(u,\bar{z})}{r^2}\,,
\ee
\begin{equation*}
\Phi_2=\frac{2(1+|z|^2)}{r}\,h(u,z,\bar{z})-\frac{(1+|z|^2)}{r^2}\,\frac{\partial Q}{\partial\bar{z}}\,.
\end{equation*}
Here, the functions $Q(u,\bar{z})$ and $h(u,z,\bar{z})$ are subject to the single field equation
\be\label{swavefe}
\frac{\partial h}{\partial z}=\frac{\dot{Q}}{(1+|z|^2)^2}\,.
\ee
Intuitively, it is clear that the characteristic data is controlled by the function $h$, which in turn determines the `generalized electromagnetic charge' $Q$ through \eqref{swavefe}. The radiative condition is simply that when integrating $\dot{Q}$ -- determined by \eqref{swavefe} -- to obtain $Q$, the function of integration $q(\bar{z})$ is set equal to zero.

Applying the double copy map (without bothering to translate the various expressions into homogeneous coordinates) gives characteristic data for the radiative gravitational and complex scalar fields
\be\label{swavegr}
\psi_4=(1+|z|^2)^2\,\int^{u}\d s\,h^{2}(s,z,\bar{z})\,, \qquad \varphi^{(0)}=(1+|z|^2)^2\,\iiint^{u}|h|^2(s,z,\bar{z})\,,
\ee
respectively. One can also generalise to the case of distinct spherical waves, determined by $h(u,z,\bar{z})$ and $f(u,z,\bar{z})$, in which case $h^2\to h\,g$, etc. in the expressions above.

This example illustrates two important properties of this classical double copy prescription. Firstly, it applies to radiative fields which are not totally null, or algebraically special of type N: the spherical wave is only totally null when $h=h(u,\bar{z})$, in which case $Q$ vanishes. Secondly, although the gauge theory input was a known exact solution, the gravitational output does not (as far as we know) match the characteristic data of any known exact solution of the Einstein equations, at least for generic $h(u,z,\bar{z})$. In other words, the double copy prescription does not necessarily map known solutions to known solutions (in contrast with the Kerr-Schild~\cite{Monteiro:2014cda} or Weyl~\cite{Luna:2018dpt} double copies, for instance).

%%%%%%%%%%%%%%%%%%%%%%%%%%

\subsection{Relation to the twistorial double copy}

Finally, we remark that this double copy prescription for characteristic data sheds some light on the `twistorial' classical double copy of~\cite{White:2020sfn,Chacon:2021wbr,Chacon:2021hfe} for linear fields. This twistorial double copy makes use of the \emph{Penrose transform}, an isomorphism between all (suitably analytic) solutions of the zero-rest-mass equations in Minkowski space-time and certain cohomology classes on \emph{twistor space} $\PT$~\cite{Penrose:1969ae,Eastwood:1981jy}:
\be\label{PTrans}
\left\{\mbox{helicity } h \mbox{ z.r.m. fields on } \M\right\}\cong H^{1}(\PT,\cO(2h-2))\,,
\ee
where $\PT$ is a (suitably chosen) open subset of $\CP^3$, and $\cO(k)$ is the sheaf of holomorphic functions of homogeneity $k$ on $\PT$.  Working with explicit `elementary states' (whose twistor representatives are rational functions~\cite{Penrose:1986uia}) in a \v{C}ech representation of $H^1(\PT,\cO(2h-2))$, the twistorial double copy for negative helicity fields takes the form
\be\label{PTDC}
f_{(-6)}=\frac{f_{(-4)}\,g_{(-4)}}{f_{(-2)}}\,,
\ee
where the subscripts denote the homogeneity weight (i.e., $f_{(-k)}\in H^{1}(\PT,\cO(-k))$). The Penrose transform ensures that this leads to a correspondence of space-time zero-rest-mass fields
\be\label{WeylDC}
\psi_{\alpha\beta\gamma\delta}=\frac{1}{\Phi}\,\phi_{(\alpha\beta}\,\phi'_{\gamma\delta)}\,,
\ee
which is equivalent to the Weyl classical double copy~\cite{Luna:2018dpt}.

At first glance, there are some problems with this prescription. The twistor representatives appearing in \eqref{PTDC} are cohomology classes, so \emph{any} representative within their equivalence class in $H^1$ will produce the same space-time field. This `twistor gauge freedom' is destroyed by this double copy prescription, since shifting the representatives on the right-hand-side of \eqref{PTDC} by exact pieces does not shift the left-hand-side by an exact term~\cite{Chacon:2021wbr}. More generally, the notion of multiplying and dividing by cohomology classes according to \eqref{PTDC} is not well-defined; this is particularly clear when using a different representation of the cohomology -- for instance, Dolbeault rather than \v{C}ech.

On the other hand, by simply treating specific \v{C}ech representatives as functions, the prescription \eqref{PTDC} does lead to non-trivial examples of the Weyl double copy~\cite{White:2020sfn,Chacon:2021wbr}. Thus, one can ask: is there a way to make this double copy prescription -- or one like it -- unambiguous? By making use of the Kirchoff-d'Adh\'emar integral formula \eqref{KdA}, our classical double copy prescription for radiative characteristic data provides an answer to this question.

\medskip

Let $Z^{A}=(\mu^{\dot\alpha},\lambda_{\alpha})$ be homogeneous coordinates on $\CP^3$, and $\PT$ be the open subset defined by $\lambda_{\alpha}\neq 0$. The non-local relationship between complexified Minkowski space and $\PT$ is captured by the incidence relations
\be\label{increl}
\mu^{\dot\alpha}=x^{\alpha\dot\alpha}\,\lambda_{\alpha}\,,
\ee
which state that a space-time point $x$ corresponds to a holomorphic complex projective line $X\cong\CP^1$ in twistor space. The Lorentzian-real slice corresponds to all lines $X$ which are completely contained in $\PN=\{Z\in\PT|Z\cdot\bar{Z}=0\}$, where
\be\label{PN}
Z\cdot\bar{Z}:=[\mu\,\bar{\lambda}]+\la\lambda\,\bar{\mu}\ra\,.
\ee
Conversely, and point $Z\in\PN$ corresponds to a null geodesic in Minkowski space.

Since every null geodesic in Minkowski space-time is uniquely identified with a point on $\scri^+$ (i.e., the point where the null geodesic intersects $\scri^+$ in the conformal compactification), there is a natural projection
\be\label{PN2}
p:\PN\to\scri^+\,, \qquad (\mu^{\dot\alpha},\,\lambda_{\alpha})\mapsto ([\mu\,\bar{\lambda}],\,\lambda_{\alpha},\,\bar{\lambda}_{\dot\alpha})\,,
\ee
with $u=[\mu\,\bar{\lambda}]$ real as a consequence of $Z\cdot\bar{Z}=0$. Thus, the twistor line $X$ is sent to the light cone cut $S_{x}^{2}\subset\scri^+$ under this projection.

With this in mind, it can be shown that characteristic data defines Penrose transform representatives~\cite{Mason:1986}. Consider the negative helicity characteristic data for a Maxwell field, $\phi_2\in\cO(-3,-1)$; the Kirchoff-d'Adh\'emar integral formula constructs the associated zero-rest-mass field on Minkowski space as:
\be\label{MaxKdA}
\begin{split}
\phi_{\alpha\beta}&=\int_{S^2_x}\lambda_{\alpha}\,\lambda_{\beta}\,\D\lambda\wedge\D\bar{\lambda}\,\left.\frac{\partial\phi_2}{\partial u}\right|_{S_x^2} \\
 &=\int_{X}\lambda_{\alpha}\,\lambda_{\beta}\,\D\lambda\wedge\D\bar{\lambda}\,\left.p^*\frac{\partial\phi_2}{\partial u}\right|_{X}\,,
\end{split}
\ee
where in the second line, we view the integrand as pulled back to $\PN$ by \eqref{PN2} and restricted to the twistor line $X$ by the incidence relations \eqref{increl}. The object $\partial_u\phi_2$ is well-defined on twistor space, since $p^{*}\phi_2$ depends on $\mu^{\dot\alpha}$ only through the combination $[\mu\,\bar{\lambda}]$.

Now, consider the object
\be\label{PTdata}
f(Z)=  \D\bar{\lambda}\,p^*\frac{\partial\phi_2}{\partial u}([\mu\,\bar{\lambda}],\lambda,\bar{\lambda})\,,
\ee
which takes values in $\Omega^{0,1}(\PT,\cO(-4))$. This obeys $\dbar f=0$, where $\dbar$ is the Dolbeault operator on $\PT$, since the only anti-holomorphic dependence in $f$ is through $\bar{\lambda}_{\dot\alpha}$ and $\D\bar{\lambda}\wedge\D\bar{\lambda}=0$. Similarly, for non-trivial data it is clear that $f\neq\dbar g$ for some $g\in\Omega^{0}(\PT,\cO(-4))$. Thus, $f$ is actually a Dolbeault cohomology class on $\PT$: $f\in H^{0,1}(\PT,\cO(-4))$. In particular, the characteristic data defines a particular representative within this cohomology class~\cite{Mason:1986}; namely, the one which takes the form \eqref{PTdata}.

Similarly, the double-copied data also defines twistor representatives within a cohomology class from the gravitational and scalar characteristic data. Indeed, these are given by
\be\label{PTdata2}
f(Z)=\D\bar{\lambda}\,p^*\tr(\phi_2^{(1)}\phi_{2}^{(2)})\in H^{0,1}(\PT,\cO(-6))\,,
\ee
for the negative helicity gravitational representative and
\be\label{PTdata3}
f(Z)=\D\bar{\lambda}\,p^*\iint^{u}\d s\,\tr(\phi_2^{(1)}\bar{\phi}_{2}^{(2)})\in H^{0,1}(\PT,\cO(-2))\,,
\ee
for the (complex) scalar representative. For the positive helicity graviton, a similar procedure can be performed to obtain a twistor representative on \emph{dual} twistor space which is the complex conjugate of \eqref{PTdata2}. Alternatively, a positive helicity cohomology representative can be constructed on twistor space itself~\cite{Mason:1986}, but requires using the triple integral of $\bar{\psi}_4$ (i.e., the integral of  $\bar{\bigma}^0$). Unfortunately, this object is not invariant under supertranslations, so such a representative actually suffers from similar problems to the original twistor double copy, in the sense that it is not gauge invariant.

%%%%%%%%%%%%%%%%%%%%%%%%%%
%%%%%%%%%%%%%%%%%%%%%%%%%%

\section{Asymptotic Classical Double Copy}
\label{AsyClassDC}

In the previous sections we presented a new notion of classical double copy, under which the radiative data at null infinity is mapped from gauge theory to gravity. By committing to purely radiative solutions, we were able to provide an elegant double copy prescription that preserves the radiative structure of the solutions.
It is an interesting question to ask whether a notion of asymptotic double copy exists more generally, when all field components (and not only the radiative ones) are present. In this section we will show that such a notion exists by studying the equations of motion order by order in the asymptotic expansion around null infinity (also known as the \emph{peeling theorem} \cite{Newman:1968uj}). We will show that order-by-order in this expansion, Maxwell's equations are mapped into Einstein's equations.

In general, this double copy prescription does not preserve the radiative structure of the solutions (as opposed to the `radiative' notion of double copy studied in the previous sections), namely it mixes the radiative degrees of freedom with the Coulombic field components.
However, in this case we are able to relate the prescription to a double copy property of scattering amplitudes.

%%%%%%%%%%%%%%%%%%%%%%%%%%

\subsection{Electrodynamics}

Let us start by reviewing the asymptotic expansion of Maxwell's equations around null infinity (see for example \cite{Strominger:2017zoo} and references therein).
We use the following convention for the action of Maxwell's fields coupled to matter
\begin{equation}
S_{\text{QED}} = \frac{1}{e^2}\int \d^4 x \sqrt{-\eta} \Big(  - \frac{1}{4} F^2  + \mL_{\text{matter}} \Big),
\end{equation}
where $\eta$ is the flat Minkowski metric.
The equations of motion that results from the action above are Maxwell's equations
\begin{equation}\label{MaxwellsEqns}
\begin{aligned}
\d * F &=  * J & \qquad \longleftrightarrow\qquad & \:\:\:\: \partial^{\nu} F_{\nu\mu} =   J_{\mu}\,, \\
\d F &= 0 & \qquad \longleftrightarrow\qquad & \:\:\:\: \pa_{[\mu} F_{\nu\rho]} = 0\,,
\end{aligned}
\end{equation}
where $F_{\mu\nu}=\pa_{\mu}A_{\nu}-\pa_{\nu}A_{\mu}$ is the field strength, the square brackets $[ \dots ]$ denote anti-symmetrization of the indices and $J$ is the matter current defined by
\begin{equation}
J _{\mu} = - \frac{\delta S_M}{\delta A^{\mu}},
\end{equation}
which is conserved if $\partial^{\mu} J_{\mu}=0$.
In retarded null coordinates $(u,r,z,\bar{z})$ Maxwell's equations become
\[
\gammaflat r^2 \pa_u F_{ur} + \pa_{(z}F_{\zb)u} -\pa_r \left(\gammaflat r^2 F_{ur}\right)
&=
 \gammaflat r^2 J_u,
\\
\pa_{(z}F_{\zb)r}  -\pa_r\left(\gammaflat r^2 F_{ur}\right)
&=
 \gammaflat r^2 J_r,
\\
r^2 \pa_r \left( F_{rz} -F_{uz} \right) - r^2 \pa_u F_{rz} -\pa_z \left(\gammaflatt F_{z\zb}\right)
&=
 r^2 J_z,
\\
r^2 \pa_r \left( F_{r\zb} -F_{u\zb} \right) - r^2 \pa_u F_{r\zb} +\pa_{\zb} \left(\gammaflatt F_{z\zb}\right)
&=
 r^2 J_{\zb},
\]
where $\gamma_{z\bar{z}}:=2\times(1+|z|^2)^{-2}$ denotes the conformal factor of the round sphere metric.

With the retarded radial gauge choice \eqref{GaugeCondition} and the asymptotic expansion \eqref{AFgauge0}, the field strength can now be expanded as
\begin{eqnarray}
F_{z\zb} &=& \left(\pa_z A_{\zb}^{(0)}  -   \pa_{\zb} A_{z}^{(0)}\right)+  \frac{\left(\pa_z A_{\zb}^{(1)}  -   \pa_{\zb} A_{z}^{(1)}\right) }{r} + O\left( r^{-2} \right)  \\
F_{u z} &=& \pa_u A_{z}^{(0)} +\frac{1}{r}\left(\pa_u A_z^{(1)} -\pa_z A_u^{(1)} \right) +O\left( r^{-2} \right)   \\
F_{r z} &=& - \frac{1}{r^2}  A_{z}^{(1)}  - \frac{2}{r^3}  A_{z}^{(2)} + O\left( r^{-4} \right)  \\
F_{ur} &=&  \frac{1}{r^2}  A_{u}^{(1)} + \frac{2}{r^3}  A_{u}^{(2)} + O\left( r^{-4} \right)
\end{eqnarray}
Assuming that $J_u$ decay as $r^{-2}$ at large distances, conservation of the current implies
\[
J_u &= \frac{j_u(u,z,\bar{z})}{r^2} + O(r^{-3}) \\
J_r &= \frac{j_r(u,z,\bar{z})}{r^4} + O(r^{-5}) \\
J_z &= \frac{j_z(u,z,\bar{z})}{r^2} + O(r^{-3}) \\
J_{\zb} &= \frac{j_{\zb}(u,z,\bar{z})}{r^2} + O(r^{-3})\,.
\]
Namely, $J_z$ and $J_{\zb}$ are of the same order as $J_u$, but $J_r$ is subleading.

\medskip

\paragraph{Asymptotic equations of motion:} At leading order in the asymptotic expansion near $\scri^+$, the $u$-component of Maxwell's equations takes the form
\[\label{uMaxEq}
\gammaflat \pa_u A_u^{(1)} = \pa_u \left(  \pa_z A_{\zb}^{(0)}  +     \pa_{\zb} A_{z}^{(0)}     \right) + \gammaflat j_{u}  .
\]
Similarly, at leading order the $z$- and $\zb$- components of Maxwell's equations are
\[\label{zMaxEq}
2\pa_u A_z^{(1)} - \pa_z A_u^{(1)} -\pa_z \left( \gammaflatt \pa_{[z}A_{\zb]}^{(0)}  \right) &=  j_z ,
\\
2\pa_u A_{\zb}^{(1)} - \pazb A_u^{(1)} +\pazb \left( \gammaflatt \pa_{[z}A_{\zb]}^{(0)}  \right) &=  j_{\zb} .
\]
The $r$-component of Maxwell's equations is a constraint equation that fixes $A_u^{(2)} $
\[\label{rMaxwell}
A_u^{(2)}  = -\frac{1}{2} \gammaflatt \pa_{(z} A^{(1)}_{\zb)} + \frac{1}{2} j_r.
\]

\medskip

\paragraph{Monopole moments and asymptotic charges:} The asymptotic charges at $\scri^+$ are defined using the leading components of the Newman-Penrose Coulomb scalar $\Phi_1$ from \eqref{NPscalars}
\[\label{NPCharges}
Q (\ep ) + \im\, \Qt (\ep) =  \frac{1}{e^2} \int _{\scri^+_-}\ep(z,\bar{z}) \, \phi_1\,,
\]
where $\scri^{+}_{-}$ is the $u\to-\infty$ limit of a constant $u$ spherical cut of $\scri^+$, $\ep(z,\bar{z})$ is the asymptotic gauge parameter and $\phi_1$ is defined by \eqref{GFpeeling} and represents the monopole moments (electric and magnetic) of the Coulomb component of the gauge field. In terms of the asymptotic expansion of the gauge field we have
\[\label{gaugeMonopole}
\phi_1 = - A_{u}^{(1)} - \gammaflatt F_{z\zb}^{(0)}\,.
\]
The electric charge at $\scri^+$ is then given by
\begin{equation}
\begin{aligned}
Q (\ep )&=
\frac{1}{e^2} \int _{\scri^+_-}\ep \; \mathrm{Re}\,   \phi_1 =
\frac{1}{e^2} \int _{\scri^+_-}\ep  *F   \\
&=\frac{1}{e^2} \int _{\scri^+_-} \d^2 z \, \gammaflat \, \ep \, F_{ru}^{(2)}
=-\frac{1}{e^2} \int _{\scri^+_-} \d^2 z \, \gammaflat \, \ep \, A_{u}^{(1)}\,,
\end{aligned}
\end{equation}
and it can be decomposed into soft and hard parts
\begin{equation}
Q (\ep)  = \Qsoft  (\ep) +\Qhard  (\ep).
\end{equation}
The soft component is
\begin{equation}
\begin{aligned}
\Qsoft  (\ep)  &= \frac{1}{e^2}  \int _{\scri^+} \d \ep \wedge *F  \\
&=
\frac{1}{e^2} \int _{\scri^+} \d u\, \d^2 z  \,  \ep \,     \pa_u \left(  \pa_z A_{\zb}^{(0)}  +     \pa_{\zb} A_{z}^{(0)}     \right)\,,
\end{aligned}
\end{equation}
and the hard component is given by
\begin{equation}
\Qhard  (\ep)  = \frac{1}{e^2}\int_{\scri^+} \ep *j =
\frac{1}{e^2}\int _{\scri^+} \d u\, \d^2 z\,  \gammaflat \,  \ep    \, j_{u} .
\end{equation}
The magnetic charge\footnote{Note that we have defined the electric and magnetic charges in \eqref{NPCharges} on equal footing. Another common definition of the magnetic charge $\Qt   (\ep) = \frac{1}{2\pi} \int_{\scri^+_-} \ep F$ differs from the one that we use by a factor of $\frac{2\pi}{e^2}$.}
\[\label{magneticCharge}
\Qt   (\ep) &=
\frac{1}{e^2} \int _{\scri^+_-}\ep \; \mathrm{Im} \,  \phi_1
 = \frac{1}{e^2} \int_{\scri^+_-} \ep\, F\\
 & = \frac{\im}{e^2}\int_{\scri^+_-} \d^2 z \, \ep \, F_{z\zb}
= \frac{\im}{e^2}\int_{\scri^+_-} \d^2 z \, \ep \,  \left(  \pa_z A_{\zb}^{(0)}  -    \pa_{\zb} A_{z}^{(0)}     \right)
\]
receives contributions from soft photons \emph{only}.
Had we included an explicit magnetic current in the Maxwell's equations \eqref{MaxwellsEqns}, the magnetic charge would have received a hard contribution as well. However, for the purpose of this paper we will not include such magnetic currents.

\medskip

\paragraph{Dipole moments:} The dipole moment of the Coulomb component of the gauge field is described by the coefficient $\phi_1^{(1)} $ \cite{Janis:1965tx,Newman:1968uj} in the expansion \eqref{GFpeeling}. In terms of the asymptotic expansion coefficients of the gauge field it reads
\[
\phi_1^{(1)} = - A_u^{(2)} - \gammaflatt F_{z\zb}^{(1)},
\]
where $F_{z\zb}^{(1)} = \pa_z A_{\zb}^{(1)}-\pazb A_{z}^{(1)}$.
Using the $r$ component of Maxwell's equations \eqref{rMaxwell}, which constraints the value of $A_u^{(2)} $ we arrive at
\[\label{gaugeDipole}
\phi_1^{(1)} = 2 D^z A_z^{(1)} - j_r\,,
\]
where we are using the covariant derivative on the sphere $D^z$, which is a different representation of the spin-weighted covariant derivative $\bar{\eth}$. Therefore, $A_z^{(1)}$ essentially describes the two dipole moments (electric and magnetic).

%%%%%%%%%%%%%%%%%%%%%%%%%%%

\subsection{Gravity}

We now turn to study the asymptotic expansion of Einstein's equations, while using the following convention for the gravitational action
\begin{equation}
S_{\text{gravity}}= \int \d^4 x \sqrt{-g} \Big(\frac{1}{16\pi G}R + \mL_{\text{matter}}\Big),
\end{equation}
where $G$ is Newton's constant.
The resulting Einstein equations are
\begin{equation}\label{EinsteinEqns}
R_{\mu\nu} - \frac{1}{2} g_{\mu\nu}R = 8\pi G \, T_{\mu\nu}^M,
\end{equation}
where the stress tensor is defined by
\begin{equation}
T^M_{\mu\nu} = -\frac{2}{\sqrt{-g}} \frac{\delta \Big(\sqrt{-g} \mL_{\text{matter}}\Big)}{\delta g^{\mu\nu}}.
\end{equation}

\medskip

\paragraph{The Bondi metric:} Similar to the electromagnetic case we will now turn to study the expansion of Einstein's equations at large distances. First we have to choose a gauge for the field (the metric in this case) and then expand. We choose to work in Bondi-Sachs gauge, in which the expansion of the metric takes the form \eqref{BS1}. Let us reproduce this expansion in Bondi-style notation, including some further subleading terms:
\begin{equation}\label{BondiMetric}
\begin{aligned}
\d s^2 &= -\d u^2 -2\,\d u\,\d r+2r^2 \gammaflat\, \d z\, \d\zb \\
&+
\frac{2\,m_B}{r}\, \d u^2 +r\, C_{zz}\, \d z^2 +r\, C_{\zb\zb}\, \d\zb ^2 -2\,U_z\, \d u\, \d z -2\, U_{\zb}\, \d u \d\zb \\
&+ \frac{1}{r}\left(
\frac{4}{3}\left(N_z+u \paz m_B \right) - \frac{1}{4} \paz \left(C_{zz}C^{zz}\right)
\right) \d u\, \d z + \mathrm{c.c.}
\\
&+ \dots ,
\end{aligned}
\end{equation}
where
\[
U_z = - \frac{1}{2} D^z C_{zz}\,.
\]
The dots in \eqref{BondiMetric} represent subleading terms in the asymptotic expansion.
In the literature $N_z$ is commonly referred to as the \emph{angular momentum aspect}, although as discussed in~\cite{Kol:2020vet} it describes both the angular momentum and the center of mass, therefore generating the entire Lorentz group.

\medskip

\paragraph{The asymptotic Einstein equations:} Plugging the expansion \eqref{BondiMetric} into the equations of motion \eqref{EinsteinEqns} we can now derive the asymptotic form of the Einstein equations for the different field components. At leading order the $uu$-component of the Einstein equations is
\[\label{uEinEq}
\pa_u m_B  =  \frac{1}{4} \pa_u  \left(D_z^2 C^{zz}+D_{\zb}^2 C^{\zb\zb}\right) -T_{uu},
\]
where
\[
T_{uu} \equiv  4\pi G \lim_{r \rightarrow \infty } \left(r^2 T^M_{uu}\right) + \frac{1}{4} N_{zz}N^{zz}
\]
and $N_{zz}=\partial_u C_{zz}$ is the news tensor in Bondi notation. The $uz$- and $u\zb$-components of the Einstein equations, at leading order in the asymptotic expansion, take the form
\[\label{zEinEq}
\pa_u N_z &= +\frac{1}{4}\paz  \left(D_z^2 C^{zz} - D_{\zb}^2 C^{\zb\zb}\right)  -u \pa_u \paz m_B  - T_{uz} ,\\
\pa_u N_{\zb} &= - \frac{1}{4}\pazb  \left(D_z^2 C^{zz} - D_{\zb}^2 C^{\zb\zb}\right)  -u \pa_u \pazb m_B  - T_{u\zb} ,\\
\]
where
\[
T_{uz} &\equiv 8 \pi G \lim_{r \rightarrow \infty } \left(r^2 T^M_{uz}\right) - \frac{1}{4} \paz \left( C_{zz}N^{zz}\right) -\frac{1}{2}C_{zz}D_z N^{zz},\\
T_{u\zb} &\equiv 8 \pi G \lim_{r \rightarrow \infty } \left(r^2 T^M_{u\zb}\right) - \frac{1}{4} \pazb \left( C_{\zb\zb}N^{\zb\zb}\right) -\frac{1}{2}C_{\zb\zb}D_{\zb} N^{\zb\zb}\,.
\]
All other components of the Einstein equations are subleading in the asymptotic expansion.

\medskip

\paragraph{Monopole moments and asymptotic charges:} As in electrodynamics, we can now express the monopole moments of the gravitational Coulomb field $\Psi_2$, defined in \eqref{NPWeyl}, in terms of the coefficients of the field's asymptotic expansion \eqref{GRpeeling}
\[\label{gravityMonopole}
\psi_2  = - m_B - \im\, \mt_B - \frac{1}{4}C_{zz}N^{zz} ,  
\]
where the dual mass aspect is given by \cite{Kol:2020zth,Kol:2020vet}
\[
\mt_B = -\frac{\im}{4} \left(  D_z^2 C^{zz}-D_{\zb}^2C^{\zb\zb} \right)\,.
\]
The leading BMS charges are then defined by
\[
T(f) +\im\, \Tt (f) = - \frac{1}{4\pi G}
\int_{\scri^+_-}  f(z,\zb) \; \psi_2\,,
\]
where $f(z,\bar{z})$ is the supertranslation parameter.
The real part defines the BMS supertranslation charge
\[
T(f) &= - \frac{1}{4\pi G}
\int_{\scri^+_-}  \d^2 z\, \gammaflat f(z,\zb)\, \mathrm{Re}\,\psi_2 \\
&= \frac{1}{4\pi G}
\int_{\scri^+_-} \d^2 z\, \gammaflat f(z,\zb)\, m_B\,, 
\]
which can be decomposed into soft and hard parts
\begin{equation}
T(f) = \Tsoft (f) + \Thard (f)\,.
\end{equation}
The hard part of the charge is given by
\begin{equation}
\Thard (f) = \frac{1}{4 \pi G} \int _{\scri^+} \d u \d^2z\, f(z,\zb)\, \gammaflat\, T_{uu}
\end{equation}
and its soft part is
\begin{equation}
\begin{aligned}\label{tsoft}
\Tsoft(f) &= \frac{1}{8\pi G} \int_{\scri^+} \d u\,  \d^2z  \, \pa_u \Big(  \pazb U_{z}  +   \paz U_{\zb}  \Big) f(z,\zb) \\
&=
- \frac{1}{16\pi G} \int_{\scri^+}  \d u \,  \d^2z \, \gammaflatt  \,  \Big(  D^2_{\zb} N_{zz}+ D^2_{z} N_{\zb \zb}    \Big)  f(z,\zb) .
\end{aligned}
\end{equation}
The dual supertranslation charge \cite{Godazgar:2018qpq,Kol:2019nkc}
 is given by
\[
\Tt(f)   &= - \frac{1}{4\pi G}
\int_{\scri^+_-}   \d^2 z\, \gammaflat\, f(z,\zb)\,    \mathrm{Im}\,\psi_2
\\
&=  \frac{\im}{16\pi G}
\int_{\scri^+_-}  \d^2 z \,  \gammaflatt\, f(z,\zb)
\Big(
D_{\zb}^2 C_{zz} -D_z^2 C_{\zb\zb}
\Big)\,,
\]
and receives contributions from soft modes only.

\medskip

\paragraph{Dipole moments:} The gravitational dipole moments are described by the first subleading term in the asymptotic expansion of $\Psi_2$:
\[\label{gravityDipole}
\psi_2^{(1)} = D^z(N_z+\paz m_B) +T_{ur}+ \text{non-linear terms}
\]
Here $T_{ur}=8\pi G \lim_{r \rightarrow \infty} \left(r^4 T_{ur}^{M}\right)$ and the precise form of the non-linear terms can be found in \cite{Kol:2020vet}.
The real and imaginary parts of $\psi_2^{(1)}$ describe the `electric' and `magnetic' dipole moments of the gravitational Coulomb field, which generate in turn boosts and rotations, respectively~\cite{Kol:2020vet}.

%%%%%%%%%%%%%%%%%%%%%%%%%%%%%%

\subsection{Double copy prescription}

It is now evident that the asymptotic form of Maxwell's evolution equations \eqref{uMaxEq}--\eqref{zMaxEq} can be mapped into the asymptotic Einstein evolution equations \eqref{uEinEq},\eqref{zEinEq} using the following replacements of the fields. First, the radiative data at null infinity is mapped as
\[\label{nullMap1}
A_z^{(0)}  \Big|_{\scri^+}    \quad& \longleftrightarrow&-&  \frac{1}{4} \pa^z C_{zz}  \Big|_{\scri^+},
\]
together with the mapping of the sources
\[\label{nullMap2}
j_u \Big|_{\scri^+}   \quad& \longleftrightarrow & T_{uu}\Big|_{\scri^+}\, , \\
j_z \Big|_{\scri^+}   \quad& \longleftrightarrow & T_{uz}\Big|_{\scri^+} \,, \\
j_{\zb} \Big|_{\scri^+} \quad& \longleftrightarrow & T_{u\zb}\Big|_{\scri^+} \,.
\]
For the rest of the field components, it is enough to make the following identification at \emph{spatial infinity}
\[\label{spatialMap}
\text{Monopole:} \qquad A_u^{(1)} \Big|_{\scri^+_-}  \quad& \longleftrightarrow&-&  m_B\Big|_{\scri^+_-}  \,,\\
\text{Dipole:} \qquad A_z^{(1)}\Big|_{\scri^+_-}  \quad& \longleftrightarrow&-&   \frac{1}{2}  \Big[ N_z + u\paz m_B \Big]_{\scri^+_-} .
%\vdots  \qquad \qquad \vdots \qquad  \quad& \longleftrightarrow&&  \vdots
\]
The map \eqref{spatialMap} at spatial infinity is enough to ensure that these field components are also mapped along the entire null boundary $\scri^+$. The reason is that only the radiative data needs to be specified on the entire null boundary and in turn it determines the development of all the other field components (for which we only need to specify boundary conditions at $\scri^+_-$) along $\scri^+$.
Note that this map is non-linear since $T_{uu},T_{uz},T_{u\zb}$ include non-linear self-interaction graviton terms.

\medskip

\paragraph{Monopole moments and asymptotic charges:} Under the asymptotic classical double copy \eqref{nullMap1}--\eqref{spatialMap}, the electric and magnetic fields are mapped into
\[
F_{ru} \quad& \longleftrightarrow&+&\frac{m_B}{r^2},\\
F_{z\zb}  \quad& \longleftrightarrow&-& \im\, \gammaflat\, \mt_B .
\]
Not surprisingly, note that $\frac{m_B}{r^2}$ is nothing but the Newtonian field, which is mapped into the electric Coulomb field $F_{ru}=-\frac{A_u^{(1)}}{r^2}$.
Perhaps more interestingly, the gravitational-magnetic field $-\im \gammaflat \mt_B $ is mapped into the magnetic field in Maxwell's theory~\cite{Kol:2019nkc,Huang:2019cja,Kol:2020ucd}.

In general, the Coulomb monopole modes \eqref{gaugeMonopole} and \eqref{gravityMonopole} do not map into each other under the asymptotic classical double copy due to the presence of a non-linear gravitational term.
However, in the absence of radiation ($N_{zz}|_{\scri^+_-}=0$) the non-linear term vanishes and
\[
\phi_1  \quad& \longleftrightarrow&  - \psi_2 \, .
\]
The asymptotic charges are defined using the Coulomb monopole modes evaluated at spatial infinity $\scri^+_-$, where we assume that there is no radiation. We therefore conclude that under the asymptotic classical double copy the asymptotic charges are mapped into each other as
\[\label{MapCharges}
Q(\epsilon) \quad& \longleftrightarrow & T(f)\,, \\
\Qt(\epsilon) \quad& \longleftrightarrow &  \Tt(f)\,,
\]
provided that we map the gauge parameter into the diffeomorphism parameter in gravity
\[
\ep  \quad& \longleftrightarrow &  f\,.
\]
Note that the soft and hard parts of the electric charge are mapped into the corresponding parts of the supertranslation charge
\[
\Qsoft (\epsilon) \quad& \longleftrightarrow &  \Tsoft (f)\,, \\
\Qhard (\epsilon)  \quad& \longleftrightarrow &  \Thard (f) \,.
\]
On the other hand, the dual supertranslation charge and the magnetic charge receive contributions from soft modes only \cite{Kol:2019nkc}.
Thus, the asymptotic double copy defined by this map preserves the hard and soft sectors of the theories.

Our result \eqref{MapCharges} generalizes the work of \cite{Huang:2019cja}, where a similar map between the asymptotic charges was derived for the Kerr-Schild class of solutions. A similar map between the conserved charges of probe particles moving in the Kerr-Schild background was recently derived in \cite{Gonzo:2021drq}. These maps also extend to the geodesic motion of the particles \cite{Huang:2019cja,Gonzo:2021drq}.

\paragraph{Dipole moments:}

The Coulomb dipole moments of the fields, given in equations \eqref{gravityDipole} and \eqref{gaugeDipole}, do not generally map into each other since the gravitational dipole moments include non-linear terms (which do not vanish even in the absence of radiation, see \cite{Kol:2020vet}). However, note that to linear order the Coulomb dipole moments do map into each other under the asymptotic classical double copy
\[
\phi_1  \quad& \longleftrightarrow&  - \psi_2 \Big|_{\text{linearized}} \, .
\]
Let us emphasize that the asymptotic classical double copy, described by equations \eqref{nullMap1}-\eqref{spatialMap}, constitute a non-linear map between solution to the equations of motion despite the fact that it does not map the Weyl scalars into each other.

%%%%%%%%%%%%%%%%%%%%%%%%%%%%%%%%
%%%%%%%%%%%%%%%%%%%%%%%%%%%%%%%%

\section{Low-Energy Double Copy for Scattering Amplitudes}
\label{Amps}

The asymptotic double copy of the previous section takes the asymptotic form of any solution to Maxwell's equations an maps them into a solution of the asymptotic Einstein equations. It is clear that this notion of classical double copy is quite different from the one we introduced for radiative fields in Sections~\ref{DCrad}, \ref{RadEx}, since radiative and non-radiative degrees of freedom are mixed by the map. While the radiative double copy is not directly related to the usual notion of double copy in scattering amplitudes, it turns out that the asymptotic double copy \emph{is} related to the double copy of amplitudes. It shares this in common with the most well-known version of classical double copy, namely Kerr-Schild double copy~\cite{Monteiro:2014cda,Luna:2015paa}, which has been linked explicitly to the amplitudes manifestation of double copy~\cite{Arkani-Hamed:2019ymq,Huang:2019cja}.

%In the previous section we showed that the asymptotic form of any solution to Maxwell's equations can be mapped into a solution of the asymptotic Einstein equations. This map is a manifestation of the \emph{classical double copy}, which was originally suggested in \cite{Monteiro:2014cda,Luna:2015paa} to relate \emph{exact} solutions of both theories albeit of a very special form - the Kerr-Schild form:
%\[
%g_{\mu\nu} = \eta_{\mu\nu}+ 2 \kappa \phi  k_{\mu}k_{\nu}.
%\]
%Here $\kappa^2=32 \pi G$ and $k_{\mu}$ is a null, geodesic, four-vector. The authors of \cite{Monteiro:2014cda} showed that a single copy gauge field that solves the Yang-Mills equations can be constructed from the Kerr-Schild metric as follows
%\[
%A_{\mu} = \phi k_{\mu}.
%\]
%Loosely speaking, the name for this map stems from the fact that the metric is composed of two "copies" of the gauge field.
%However, this notion of the double copy is directly related to its original notion in terms of scattering amplitudes \cite{Arkani-Hamed:2019ymq,Huang:2019cja}.

Using on-shell methods, \cite{Arkani-Hamed:2017jhn} defines the a `minimally coupled' 3-point amplitude for massive particle of mass $m$ and spin $S$ coupled to photons or gravitons:
\[\label{DCamplitude}
\mM = g \left(x m \right)^h \frac{\langle \mathbf{1}\mathbf{2}\rangle^{2S}}{m^{2S}}
\]
(see figure \ref{fig:DCamplitude}).
Here $g=(\sqrt{2}e,\sqrt{8 \pi G})$ and $h=(1,2)$ for photons and gravitons, respectively.
The $x$-factor and $\langle \mathbf{1}\mathbf{2}\rangle$ are defined in terms of the kinematic data using spinor-helicity variables~\cite{Arkani-Hamed:2017jhn}; the precise definitions are not important here. This amplitude explicitly manifests the double copy structure and is directly related to the Kerr-Schild double copy~\cite{Arkani-Hamed:2019ymq,Huang:2019cja,Emond:2020lwi}.
We now show that the asymptotic double copy introduced in the previous section is directly related to the double copy of this 3-point scattering amplitude.

\begin{figure}[t]
	\centering
	\begin{tikzpicture}
	\begin{feynman}
	\vertex (li);
	\vertex [above=2cm of li] (a);
	\vertex [below left=1cm and 2cm of li] (b);
	\vertex [below right=1cm and 2cm of li] (c);
	\node[left=3cm of li,label=left:{$\mM \quad = \quad g(xm)^h \frac{\langle \mathbf{1}\mathbf{2}\rangle^{2S}}{m^{2S}} \quad$}] (A)   {$=$};

	\diagram* {
		(li) -- [photon, edge label=\(h \), momentum'=\(k\)] (a),
		(b) -- [plain, momentum=\(p\)] (li),
		(li) -- [plain, momentum=\(p-k\)] (c),
	};
	\end{feynman}
	\end{tikzpicture}
	\caption{The 3-point scattering amplitude that describes the interaction between photons or gravitons with momentum $k$ and matter fields. Here $g=(\sqrt{2}e,\sqrt{8 \pi G})$ is the coupling constant, $h=(1,2)$ is the massless particle spin and $m$ is the mass of the matter particle.}
	\label{fig:DCamplitude}
\end{figure}

%In the previous section we have shown that the classical double copy holds, in fact, more generally for any asymptotically flat solutions at large distances (or small momentum). In this section we will show that, similarly to the Kerr-Schild double copy, the asymptotic classical double copy is directly related to the double copy structure of scattering amplitudes.

\medskip

To begin with, note that under the asymptotic classical double copy the variation of the electromagnetic action
\[
\delta S_{\text{asy}}^{\text{QED}}= \frac{1}{e^2} \int \d^4x \sqrt{-\eta} \times \left(\nabla^{\nu}F_{\mu\nu}- J_{\mu}\right)_{\text{asy}} \delta A^{\mu}_{\text{asy}}
\]
maps into the variation of the gravitational action
\[
\delta S_{\text{asy}}^{\text{Gravity}}= \int \d^4x \sqrt{-\eta} \times \left(\frac{1}{16 \pi G}G_{\mu\nu}-\frac{1}{2} T_{\mu\nu}^M \right)_{\text{asy}}
\times \sqrt{32 \pi G}  \,  \delta   h^{\mu\nu}_{\text{asy}} .
\]
where $g_{\mu\nu}=\eta_{\mu\nu}+\frac{1}{\sqrt{32\pi G}}h_{\mu\nu}$ is the linearized graviton and its inverse is $g^{\mu\nu}=\eta^{\mu\nu}-\sqrt{32\pi G}h^{\mu\nu}$. Here the subscript `asy' indicates that we work to leading order in the asymptotic expansion of the fields.
Under the asymptotic classical double copy, we therefore have that
\[
\delta S_{\text{asy}}^{\text{QED}}
\quad  & \longleftrightarrow &
\delta S_{\text{asy}}^{\text{Gravity}}
\]
since the equations of motion and the fields are mapped into each other.
In particular, the asymptotic interactions with matter are mapped as
\[\label{InteractionTerms}
\mL_{\text{Asymp. Int.}}^{\text{QED}} = \frac{1}{e^2} A^{\mu}J_{\mu} \Big|_{\text{asy}}
\quad  & \longleftrightarrow &
\mL_{\text{Asymp. Int.}}^{\text{Gravity}} = \sqrt{8\pi G} h^{\mu\nu}T_{\mu\nu}^M\Big|_{\text{asy}}\,.
\]
The map takes the 3-point vertex between the electromagnetic field and matter to the 3-point vertex between the gravitational field and matter.

For a complex scalar field, coupled to the Maxwell Lagrangian, the current is given by
\[
J_{\mu}=\frac{\im}{\sqrt{2}}  e  \left( \phi\, \pa_{\mu} \bar{\phi}-\bar{\phi}\, \pa_{\mu} \phi \right),
\]
where we have normalized $J_{\mu}$ to match the conventions of \cite{Arkani-Hamed:2017jhn,Arkani-Hamed:2019ymq} for the coupling constant.
In momentum space we then have
\[
J_{\mu}=  \sqrt{2}e^2 \, p_{\mu}\,,
\]
where $p_{\mu}$ is the scalar's momentum. At leading order in the photon's momentum $k$, the 3-point amplitude describing the interaction between the scalar and a photon is therefore
\[\label{QEDamp}
\mM_{\text{QED}}=\sqrt{2} e \left( \ep ^{\mu} p_{\mu} \right) \,,
\]
where $\ep^{\mu}$ is the polarization vector of the photon. At leading order in small $k$ this result is, in fact, valid for matter particles of \emph{any} spin \cite{Weinberg:1965nx}. In other words, we see that the $x$-factor for this coupling is given by
\[
x_{\text{QED}}= \ep \cdot u\,,
\]
where $u_{\mu} = \frac{1}{m}p_{\mu}$ is the four-velocity (see~\cite{Moynihan:2020gxj} for a related discussion).

In gravity, the stress tensor of a complex scalar field is
\[
T_{\mu\nu}=\pa_{\mu}\phi\, \pa_{\nu} \bar{\phi} -\frac{1}{2}\eta_{\mu\nu} \left(\pa_{\mu}\phi\, \pa^{\mu}\bar{\phi} - m^2 |\phi|^2 \right)\,.
\]
In momentum space this gives
\[\label{stressTensor}
T_{\mu\nu}= p_{\mu}p_{\nu}-\frac{1}{2}\eta_{\mu\nu} \left(p^2 - m^2 \right),
\]
where the second term vanishes on-shell.
The 3-point amplitude that results from the interaction \eqref{InteractionTerms} is then given, in the limit of small graviton momentum $k$, by
\[\label{gravityAmplitude}
 \mM_{\text{Gravity}}= \sqrt{8\pi G} \, \ep^{\mu\nu}p_{\mu}p_{\nu}.
\]
The transverse-traceless components of the polarization tensor obey
\[\label{TTcomponents}
\ep^{\mu\nu}k_{\nu}=0 , \qquad \qquad  \ep^{\mu\nu}\eta_{\mu\nu}=0
\]
(hence the second term in the stress tensor \eqref{stressTensor} do not contribute to the amplitude \eqref{gravityAmplitude}) and can be decomposed in terms of two vectors
\[
\ep_{\mu\nu}=\ep_{\mu}\ep_{\nu},
\]
therefore allowing us to write the 3-point amplitude as
\[\label{xGravity}
\mM_{\text{Gravity}}=\sqrt{8\pi G }  \left(\ep ^{\mu} p_{\mu}\right)^2.
\]
Here, again, we worked out the example of a spin zero matter field, but in the small graviton momentum limit the result \eqref{xGravity} is valid for any spin~\cite{Weinberg:1965nx}.
In other words, the $x$-factor for this coupling is
\[
x_{\text{Gravity}}= \left( \ep \cdot u \right )^2\,.
\]
We therefore conclude that
\[
x_{\text{Gravity}}=x_{\text{QED}}^2\,,
\]
and the low-energy 3-point amplitude obeys the double-copy structure \eqref{DCamplitude}.

Our definition of the $x$-factor coincides with that of~\cite{Arkani-Hamed:2017jhn} for a scalar field. However, the $x$-factor of \cite{Arkani-Hamed:2017jhn} is defined more generally for particles of any spin using spinor-helicity variables. Here, we are only interested in the low-energy limit (of the photon/graviton) and therefore do not need to use the full machinery of~\cite{Arkani-Hamed:2017jhn} since the results in this limit are universal and independent of spin.

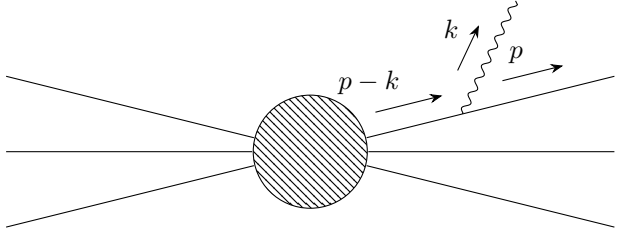
\begin{figure}[t]
	\centering
	\begin{tikzpicture}[font=\small]
	\begin{feynman}
	\vertex [blob,scale=2](li){};
	\vertex [above right = 1cm and 4cm of li] (a1);
	\vertex [above right = 0.5cm and 2cm of li] (a2);
	\vertex [above right = 1.5cm and 0.7cm   of a2] (a3);
	\vertex [right = 4cm of li] (b1);
	\vertex [below right = 1cm and 4cm of li] (c1);

	\vertex [above left = 1cm and 4cm of li] (a4);
	\vertex [left = 4cm of li] (b2);
	\vertex [below left = 1cm and 4cm of li] (c2);

	\diagram* {
		(li) -- [plain, momentum=\({ p-k}\)] (a2),
		(a2) -- [plain, momentum={[arrow shorten=0.3]\(p\)}] (a1),
		(a2) -- [photon, momentum={[arrow shorten=0.3] \(k\)}] (a3),
		(li) -- [plain] (b1),
		(li) -- [plain] (c1),
		(li) -- [plain] (a4),
		(li) -- [plain] (b2),
		(li) -- [plain] (c2),

	};
	\end{feynman}
	\end{tikzpicture}
	\caption{A Feynman diagram with a soft external photon/graviton leg of momentum $k$.}
	\label{fig:softTheorem}
\end{figure}

\medskip

Let us note that the low-energy double copy has long been known through its manifestation in soft theorems~\cite{Weinberg:1965nx}. An amplitude with a soft photon of momentum $k$ will factorize has
\[
\mA_{k} = \im\, \mM_{\text{QED}} \times \frac{-\im}{p\cdot k } \times \mA\,,
\]
(see figure \ref{fig:softTheorem}).
Here $\mA$ is the same amplitude without the external soft photon, $\mM_{\text{QED}}$ is the vertex in the small momentum limit given by \eqref{QEDamp} and the extra matter field propagator is given by $\frac{-\im}{p\cdot k }$. One then arrives at the famous soft factorization:
\[
\mA_{k} = \frac{\sqrt{2} e\, ( p \cdot \ep) }{p\cdot k } \times \mA \,.
\]
Similarly, an amplitude with an external soft graviton of momentum $k$ will factorize as
\[
\mA_{k} = \im\, \mM_{\text{Gravity}} \times \frac{-\im}{p\cdot k } \times \mA\,,
\]
with the same propagator but with a different `double-copied' vertex, to give
\[
\mA_{k} = \frac{\sqrt{8 \pi G}\,  ( p \cdot \ep)^2 }{p\cdot k } \times \mA\, .
\]
Of course, this realization of the double copy is very old news. However, it is reassuring to see that it connects naturally with the asymptotic classical double copy introduced in the previous section. Finally, let us note that the amplitude associated with the soft factors was also realized as the eikonal amplitude for gravitational Bremsstrahlung in~\cite{Luna:2016due} and was shown to obey the double copy structure.

%%%%%%%%%%%%%%%%%%%%%%%%%%%
%%%%%%%%%%%%%%%%%%%%%%%%%%%

\section{Concluding Remarks}

The classical double copy describes a set of maps between classical solutions of Yang-Mills theory and those of Einstein's theory, and is closely related to the notion of double copy in scattering amplitudes.
In this paper we have used the constraints implied by asymptotic flatness and the structure of null infinity to derive two complementary notions of the classical double copy for asymptotically flat fields.

First we considered purely radiative solutions and their characteristic data at null infinity; this data is a free function of the appropriate spin and conformal weight (or in the homogeneous formalism, a function of appropriate homogeneity). Using the radiative condition, we showed that the characteristic data of a radiative gauge field can be mapped into the characteristic data for radiative NS-NS gravitational fields. For the metric, this is described by \eqref{DCmetric}, which manifestly constructs the gravitational data as two copies of gauge theory data. In addition, this double copy prescription has the appealing property of preserving the radiative structure of the solutions: it maps the radiative components of the fields ($\phi_2 \rightarrow  \psi_4$) by construction. However, it only applies to radiative solutions, and only acts at the level of the characteristic data (rather than the fields themselves).

The second map that we derived applies more generally for any asymptotically flat solutions, but it does not preserve their radiative structure (in other words, in general the radiative and Coulomb components of the fields will mix under this map). We constructed this map by expanding the equations of motion order-by-order around null infinity. We then showed that in the first two leading orders, Maxwell's equations can be mapped into Einstein's equations. This map is directly related to the double copy structure of the 3-point amplitude which describes the leading order interaction between matter fields and photons/gravitons. This amplitude is manifested in Weinberg's soft theorems and its universality is implied by the asymptotic symmetry group. We conclude by emphasizing the important role of the asymptotic symmetry group, which not only constrains the form of classical solutions and scattering amplitudes, but also implies a concrete double copy structure.

Let us end with a couple of suggestion for future work. First, it would be interesting to understand the relationship between the asymptotic double copy prescriptions that we presented in this paper and the asymptotic Weyl double copy of \cite{Godazgar:2021} for type D and N vacuum solutions.
It would also be interesting to explore whether the low-energy double copy for scattering amplitudes that we discussed in section \ref{Amps} can be extended beyond the leading order interaction and to study potential connections to subleading soft-theorems.

\acknowledgments

TA is supported by a Royal Society University Research Fellowship. We thank the authors of~\cite{Godazgar:2021} for discussions and sharing their work prior to publication. We also thank Kevin Nguyen and Donal O'Connell for interesting comments and conversations.

%%%%%%%%%%%%%%%%%%%%%%%%%%
%%%%%% Bibliography %%%%%%
%%%%%%%%%%%%%%%%%%%%%%%%%%

\bibliography{cdc}
\bibliographystyle{JHEP}

\end{document}